\documentclass{ifacconf}
\usepackage{graphicx}      % include this line if your document contains figures
\usepackage{subcaption}     % 子图排版包（支持子标题）
\usepackage{natbib}
\usepackage{amsmath,amssymb}
\usepackage{mathtools}
\usepackage{algorithm}  
\usepackage{algpseudocode}
\usepackage{xcolor}

\usepackage{tikz}
\usetikzlibrary{decorations.pathmorphing, positioning, arrows.meta, calc, shapes.misc}

\begin{document}
\begin{frontmatter}

% 0.95 表示水平压缩到 95%
\title{\scalebox{0.95}{Distributed Articulation Point Identification} \\ \scalebox{0.95}{in Time-Varying Undirected Networks}\thanksref{footnoteinfo}}

\thanks[footnoteinfo]{
This work was supported by the National Natural Science Foundation of China under Grants U23B2060 and 62173294, and the Zhejiang Provincial Natural Science Foundation of China under Grant LZ24F030001. 
(\textit{Corresponding author: Ronghao Zheng.})
}

\author[First,Second]{Xinye Xie} 
\author[First,Second]{Ronghao Zheng} 
\author[First,Second]{Senlin Zhang}
\author[Second,Third]{Meiqin Liu}

\address[First]{College of Electrical Engineering, Zhejiang University, Hangzhou 310027, China}
\address[Second]{State Key Laboratory of Industrial Control Technology, Zhejiang University, Hangzhou 310027, China}
\address[Third]{National Key Laboratory of Human-Machine Hybrid Augmented Intelligence, Xi'an Jiaotong University, Xi'an 710049, China \\ e-mail: {\{xinye\_xie, rzheng, slzhang, liumeiqin\}@zju.edu.cn}}

\begin{abstract}                % Abstract of 50--100 words
%% 首先开门见山提出解决什么问题（识别动态网络中的割点、双连通分量、双连通性）；
% 然后讲这个问题具体实现了什么（节点本地判断自身是否为割点并与其他节点达成网络中的割点共识）；再讲实现的算法基于最大共识协议、局部含标签的更新协议，在面对网络变化时（添边减边、添节点删节点），局部更新本地信息，无需重新完全初始化本地信息。算法相较于完全初始化本地信息的方法减少了收敛轮次。
% 感觉不对，是需要强调时变后信息的更新规则，割点的识别依赖于收敛后的该信息。
% 先写具体内容再考虑摘要吧
Identifying articulation points (APs) is fundamental to assessing the robustness of time-varying networks. In such dynamic environments, topological changes including edge additions and deletions can instantly alter the set of APs, demanding rapid and efficient re-assessment. This paper proposes a fully distributed algorithm for identifying APs and monitoring biconnectivity. Our core contribution is an incremental update protocol. 
Unlike static methods that require global re-initialization which incurs high communication overhead, 
our algorithm propagates information from the site of the change, updating only the affected nodes' state values. This approach, which builds upon a maximum consensus protocol, not only ensures convergence to the correct AP set following topological changes but also preserves network privacy by preventing nodes from reconstructing the global topology. We provide rigorous proofs of correctness for this eventual convergence and demonstrate its applicability and efficiency through experiments.
% its efficiency gains over static re-computation through simulation.

% This paper addresses the problem of dynamically identifying articulation points and biconnected components and assessing the network biconnectivity in time-varying networks. A general procedure is outlined to enable individual nodes to locally determine their status as articulation points and reach consensus on their determination of articulation points in the network. Our implementation of the procedure exploits the maximum consensus protocol, yielding a distributed algorithm that updates only localized, partial state in response to edge addition or deletion, avoiding the need for global re-initialization of all the nodes' states.
% The proposed algorithm preserves network privacy of the overall network topology without global topological information.
% We provide rigorous proofs of correctness under dynamics for the proposed algorithm and demonstrate its applicability and efficiency through experiments.
\end{abstract}


\end{frontmatter}
%===============================================================================

\section{Introduction}
%网络鲁棒性和联通性。
%现存工作：中心式算法依赖全局信息。分布式计算让本地节点计算全局的割顶点，但是基于扩展树，能恢复出全局拓扑。
%引出我们的工作。贡献
\subsection{Motivation and Literature Review}
\begin{figure}[t]
    \centering  
    \includegraphics[width=0.5\linewidth]{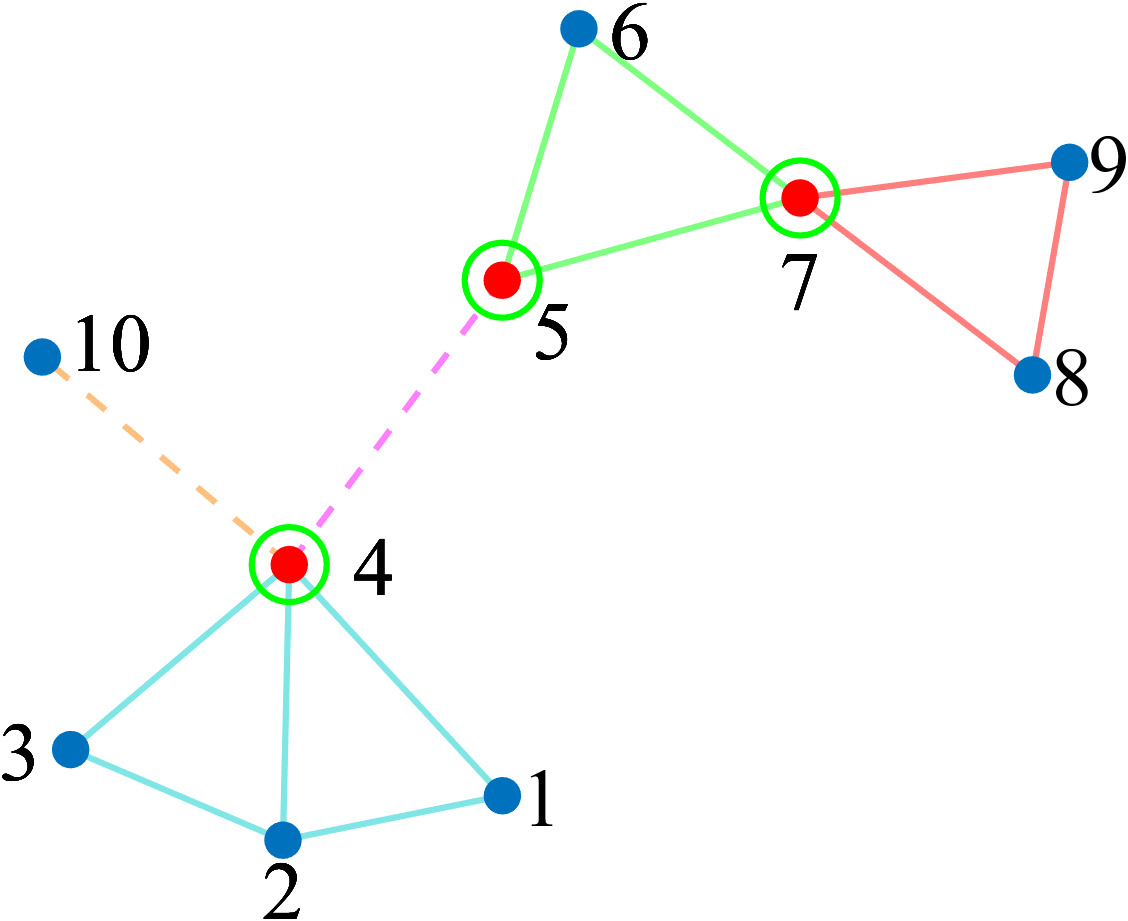}
    \caption{An undirected graph with $10$ nodes. Red nodes with green circles represent APs, namely nodes~$4, 5$, and $7$. Dashed lines represent bridges, namely edges~$(4,5)$ and $(4,10)$. Edges in different biconnected components are depicted in different colors.}
    \label{Fig:easy}
    % \vspace{-4mm}
\end{figure}

Analyzing and enhancing network robustness is crucial for ensuring the reliability of interconnected systems. A key aspect of robustness lies in a network’s ability to withstand failures while maintaining overall connectivity.
The failure of certain critical nodes, known as \emph{articulation points} (APs) or \emph{cut vertices}, can fragment a network into disconnected components (see Fig.~\ref{Fig:easy}). Therefore, \emph{biconnectivity}, the property of a network to remain connected even after the removal of any single node, is a vital robustness metric. The identification of APs is the means for assessing network biconnectivity, providing an essential foundation for enhancing robustness in systems ranging from multi-robot networks \citep{BiconnectivityRestoration,Restrepo2024CDC} to power grids \citep{Nature,PowerGrid} and social networks \citep{SocialNetworks,Access}.

The challenge of AP identification is amplified in time-varying networks, where edge additions and deletions are commonplace. These topological changes can instantly alter the set of APs. Consequently, network biconnectivity cannot be determined by a one-time calculation but must be continuously tracked.

Unfortunately, most existing algorithms for finding APs were designed for static graphs. Centralized algorithms, such as Tarjan's DFS-based method \citep{Tarjan1972,Tarjan1984}, are efficient but require global topology information, rendering them unsuitable for distributed networks. Early distributed approaches \citep{DDFS1989,CAM2012,DENCUT2014} attempted to solve this by building global structures like spanning trees. While functional, these methods have two critical flaws in the context of dynamic networks. First, they often expose the global network structure and violate privacy. Second, their only recourse to a topological change is a global re-initialization, which is typically requiring time proportional to the network size \citep{DDFS1989}. Consequently, relying on global re-initialization renders the network's information perpetually out-of-date in dynamic environments.

A more recent class of privacy-preserving methods employs the maximum consensus protocol \citep{Max-consensus} to gather necessary topological information, such as reachability or distances, without reconstructing the full graph \citep{AtmanCDC,ReedTAC,ramos2024}. This line of work is highly relevant as it respects the privacy of individual nodes. For example, \cite{venkateswaran2024CDC} successfully leverages this approach to detect bridges (edges whose removal disconnects the undirected networks) using only local distance information. However, current research in this domain focuses primarily on static networks.
Applying max-consensus protocol to time-varying networks requires mechanisms for incremental updates to avoid re-computation. Furthermore, the scope of AP identification is broader than bridge detection. While bridge endpoints are inherently APs, APs can also exist independent of any bridge (e.g., node 7 in Fig.~\ref{Fig:easy}).
%, necessitating a more general detection strategy.
% Furthermore, bridge-identification methods can only identify APs at bridge endpoints but not all APs are necessarily located at bridge endpoints. As shown in Fig.~\ref{Fig:easy}, node $7$ is an AP independent of any bridge.

While research on dynamic graph algorithms is active, it has primarily focused on centralized data structures for properties like biconnectivity \citep{Holm2001,Holm2025FullyDB} or related topological problems such as strongly connected components in directed graphs \citep{KarczmarzS24}. In contrast, distributed and dynamic solutions are scarce. Therefore, a solution that combines distributed, privacy-preserving, and incremental updates for the specific problem of AP identification remains a significant gap.
% Existing distributed approaches for dynamic graphs (e.g., \cite{Costa2017,Koutis2016,Zhang2018_Dynamic}) often focus on maintenance of biconnectivity by proactively adding edges rather than detecting APs, or they do not meet the privacy-preserving criteria of consensus-based methods. 

% Therefore, a significant research gap persists for a solution that is simultaneously distributed, privacy-preserving, and incrementally updatable. This paper bridges that gap. We propose a novel, event-driven algorithm that operates on the max-consensus framework. Instead of global re-initialization, a change (e.g., edge deletion) generates a localized ''flag'' that propagates from the event's location and triggers local state recalculations only in the nodes affected. This allows the network to efficiently and dynamically converge to the new, stable state values, which nodes use to locally identify the correct set of APs.

\subsection{Statement of Contributions}
This work builds upon and extends our preliminary study \citep{Xie2025CDC}. While \cite{Xie2025CDC} established the distributed criterion for identifying APs in static networks, this paper introduces a novel dynamic distributed algorithm for AP identification and biconnectivity check in time-varying, undirected networks. Our algorithm builds upon the privacy-preserving max-consensus framework but introduces an incremental update protocol as a non-trivial extension.

Instead of global re-initialization, a change (e.g., edge deletion) generates a localized ``flag'' that propagates from the event's location. This flag triggers localized recalculations of network state, such as reachability and distance metrics, only in the affected nodes, thereby allowing the network to converge to new, stable state values that nodes use to locally identify the correct set of APs.
% leading to a new, stable consensus.

The main contributions of this paper are twofold:

\begin{itemize} 
\item We propose a distributed algorithm for AP identification in time-varying networks. The algorithm handles edge additions and deletions through localized, incremental updates, avoiding global re-initialization and reducing convergence overhead. 
\item We establish rigorous theoretical guarantees for eventual convergence to the correct AP set after any arbitrary sequence of topological changes, assuming a period of stability. 
% \item The algorithm is privacy-preserving, inheriting the benefits of the max-consensus protocol, which prevents any individual node from reconstructing the overall network topology. 
\end{itemize}

\section{Problem Formulation}
%(首先引入标注和图理论；然后引理介绍已有分布式算法能获得的信息；引出建立在图上的网络问题，提出假设）
In this section, we establish the graph theoretic foundations and notation. Note that \emph{graph} and \emph{network} are used interchangeably hereafter. Subsequently, we formalize the dynamic AP identification problem and review the necessary preliminaries on distributed estimation.
% In this section, we review some graph concepts and we use the terms \emph{graph} and \emph{network} interchangeably throughout this paper.
% Subsequently, we formalize the problem addressed in this work.

\subsection{Notation and Graph Theory}
An \emph{undirected graph} is denoted by $\mathcal{G}=(\mathcal{V}, \mathcal{E})$ with a set of nodes $\mathcal{V}=\{1,2,\dots,n\}$ and a set of edges $\mathcal{E}\subseteq \mathcal{V}\times \mathcal{V}$. 
The presence of an edge $(i,j)\in \mathcal{E}$ denotes that information can be mutually exchanged between $i$ and $j$. The neighborhood set of node $i$ is defined as $\mathcal{N}_i = \{j \in \mathcal{V}\mid (i,j)\in \mathcal{E}\}$. 

A \emph{path} between two nodes $i_1$ and $i_p$ is a sequence of distinct nodes $(i_1, i_2\dots, i_p)$ where $p>1$ and each consecutive pair $(i_q,i_{q+1}) \in \mathcal{E}$. 
%\orange{Given a path $S$, the node set containing all distinct nodes traversed by the path $S$ is denoted by $V_S$.}
% The shortest-path distance vector of node $i$ is defined as $d_i = [d_{i1},d_{i2},\dots,d_{in}]$, where $d_{ij}$ is the length of the shortest path from node $i$ to node $j$.
The \emph{shortest-path distance} between two nodes is the length of the shortest path connecting them, measured by the number of edges in the path.
% The \emph{shortest-path distance} is the the number of edges in the shortest path between two nodes.
% $i$ and node $j$, denoted by $d_{ij}$.
A \emph{cycle} is a sequence of nodes $(i_1, i_2\dots, i_{p},i_1)$ where $p\geq 3$ and all nodes $i_1,i_2,\dots i_{p}$ are distinct.
% A \emph{cycle} is a sequence of nodes $(i_1, i_2\dots, i_{p},i_1),p\geq 3$ with at least three distinct nodes such that $i_1,i_2,\dots i_{p}$ are all distinct.
%Two paths between $i$ and $j$ are \emph{vertex-disjoint} if they have no nodes in common except for the endpoints $i$ and $j$.
% \emph{Vertex-disjoint} paths are paths sharing no intermediate nodes except for the endpoints.
A graph $\mathcal{G}$ is \emph{connected} if there exists at least a path between any pair of distinct nodes. 
%{The neighboring structure of a node consists of its neighbors and their minimum distances, illustrating how nodes are interconnected within the graph}. 
An \emph{AP} is a node whose removal results in the disconnection of the graph.
The \emph{vertex connectivity} of a graph is the minimum number of nodes that must be removed to disconnect it, indicating its robustness to node failures. A graph with vertex connectivity $1$ has at least one AP, while a \emph{biconnected} graph with vertex connectivity $2$ is robust in that it contains no APs and remains connected unless at least two nodes are removed.
\emph{Biconnected components} are maximal subgraphs in which every pair of nodes remains connected after removing any single node.
% 去掉了multiple node 和 higher vertex connectivity的描述

Within this paper, let $\mathbb{R}$ be the set of real numbers and $\mathbb{N}$ be the set of non-negative integers. 
%By 1n 2 Rn and 0n 2 Rn, we denote the all ones vector and zeros vector in n-dimension, respectively. 
For a given set $\mathcal{N}$, $|\mathcal{N}|$ denotes the number of elements in this set. The notation $\binom{\mathcal{N}}{2}$ represents the set of all unordered pairs of distinct elements from $\mathcal{N}$.
% Two sets are said to be \emph{disjoint sets} if they have no element in common.
The state associated with node $i$ is represented using subscript notation. For example, the state $a\in \mathbb{R}^n$ for node $i$ writes as $a_i$. The $j$-th element of the state $a_i$ is denoted by $a_{ij}$.

\subsection{Problem Settings} \label{Problem Settings}
Consider a time-varying undirected network $\mathcal{G}[t] = (\mathcal{V}, \mathcal{E}[t])$, where $\mathcal{V}$ is the fixed set of nodes and $\mathcal{E}[t]$ denotes the set of active edges at discrete time $t \in \mathbb{N}$. At each synchronous time $t$, immediate neighbors of node $i$ are denoted by $\mathcal{N}_i[t] = \{j \in \mathcal{V} \mid (i,j) \in \mathcal{E}[t]\}$, and the inclusive neighbor set is denoted by $\overline{\mathcal{N}}_i[t] = \mathcal{N}_i[t] \cup \{i\}$.
Each node $i$ is aware of its own unique ID, the total number of nodes $n$\footnote{The exact total number of nodes $n$ is not strictly necessary. A value $\overline{n}$ exceeding the network diameter suffices to guarantee the termination and correctness of the distributed consensus protocols. Similar parameter relaxation is discussed in \cite{Atman2024ECC}},
and at any synchronous time $t$, it has access to the states of its neighbors from the previous time step $t-1$, denoted by $\mathcal{N}_i[t-1]$.
% \orange{Throughout this paper, the terms \emph{network} and \emph{graph} are used interchangeably.}

Within this framework, this paper's objective is to develop a privacy-preserving distributed method that enables each node to identify whether it is an AP in the current graph $\mathcal{G}[t]$,
%classify its neighbors into their respective biconnected components, 
and re-assess the overall biconnectivity of the network after topological changes.

\subsection{Preliminaries on Distributed Estimation}
\label{sec:preliminaries}

To facilitate the subsequent analysis of time-varying topologies, we first establish the foundational distributed estimation protocols for static networks.
We introduce the binary reachability state $x_i[t]=\big[x_{i1}[t],\dots, \allowbreak x_{in}[t]\big]^\top \in\mathbb{R}^n$ for checking if node $i$ is reachable from other nodes ($1$ for reachable, $0$ for unreachable), and the shortest-path distance state $d_i[t]=\big[d_{i1}[t],\dots,d_{in}[t]\big]^\top \in\mathbb{R}^n$ for estimating the distance from other nodes.

When the network is initialized at $t=0$, nodes compute their states iteratively. Each node initializes its reachability state $x_i[0]$ and distance state $d_i[0]$ as
\begin{equation*}\label{initial values}
    x_{ik}[0]=\left\{\begin{array}{ll}
    1, & \text{if $k=i$},\\ [4pt]
    0, & \text{otherwise},
    \end{array}\right.
    \quad
    d_{ik}[0]=\left\{\begin{array}{ll}
    0, & \text{if $k=i$}, \\ [4pt]
    \infty, & \text{otherwise},
    \end{array}\right.
\end{equation*}
and iteratively updates them via the standard max-consensus protocol at time $t$:
\begin{equation}\label{eq:reachability}
    x_{ik}[t]=\max_{j\in \overline{\mathcal{N}}_i[t]} x_{jk}[t-1],
\end{equation}
\begin{equation}\label{eq:distance}
    d_{ik}[t]=\left\{\begin{array}{ll}
    d_{ik}[t-1], & \text{if $x_{ik}[t]=x_{ik}[t-1]$}, \\ [4pt]
    \min_{j\in \mathcal{N}_i[t-1]}(d_{jk}[t-1]+1) , & \text {if $x_{ik}[t]>x_{ik}[t-1]$}.
    \end{array}\right.
\end{equation}
If the connected network with diameter $D$ remains static for $n$ iterations, the state $x_{ik}$ converges to $1$ and $d_{ik}$ converges to the true shortest-path distance within $D$ iterations \citep{venkateswaran2024CDC}.

Since the reachability state is binary, the maximum operation is functionally equivalent to the logical disjunction (Boolean OR). Consequently, the protocol \eqref{eq:reachability} can be reformulated using the element-wise logical disjunction operator $\bigvee$ as
\begin{equation}\label{eq:standard-max-consensus}
x_i[t] = \bigvee_{j \in \overline{\mathcal{N}}_i[t]} x_j[t-1].
\end{equation}

\section{Incremental Update Mechanism for Time-Varying Networks}
\label{Section:Update Mechanisms}
The identification of APs relies on nodes possessing correct reachability and distance states \citep{Xie2025CDC}. This section presents a distributed algorithm to maintain these states through an incremental update mechanism. We adopt a update-correct architecture: nodes first estimate an intermediate state based on information from neighbors (Consensus), and subsequently adjust this state if a local edge change breaks the shortest path (Correction). This ensures that the final state preserves reachability when the modified edge was not part of the shortest path, while resetting the state if the change invalidates the current shortest path.

\subsection{Change Flag Propagation}
\label{sec:propagation}
To facilitate the incremental update, we define the change flag to represent topological events. A change flag is defined as a tuple $\phi \triangleq (\text{type}, u, v, \tau),$ where $\text{type} \in \{\text{ADD}, \text{DELETE}\}$ indicates the event type, $(u, v)$ denotes the incident nodes of the modified edge, and $\tau$ represents the generation timestamp. We treat $(u, v)$ as an unordered pair, such that $(\text{type}, u, v) \equiv (\text{type}, v, u)$.

Each node $i$ maintains a set of active change flags at time $t$, whether locally generated or received from neighbors, denoted by $\Phi_i[t]$. Let $\mathcal{H}_i[t]$ denote the historical set of all change flags encountered by node $i$ up to time $t$, initialized as $\mathcal{H}_i[0] = \emptyset$.

At time $t$, node $i$ receives flag sets from its neighbors. The set of newly received flags, denoted as $\widetilde{\Phi}_i[t]$, is derived by excluding any flags that have already been processed:
\begin{equation}\label{eq:phi_tilde_update}
    \widetilde{\Phi}_i[t] = \bigcup_{j \in \mathcal{N}_i[t-1]} \Phi_j[t-1] \setminus \mathcal{H}_i[t-1].
\end{equation}

The active flag set $\Phi_i[t]$ is initialized with these newly received flags:
\begin{equation*}
\Phi_i[t] \leftarrow \widetilde{\Phi}_i[t].
\end{equation*}
Note that this set will be further expanded to include locally generated flags in the subsequent correction step (Section~\ref{Change Detection and Local State Update}).

The node updates its history set to include these newly received flags:
\begin{equation}\label{eq:history_update}
    \mathcal{H}_i[t] = \mathcal{H}_i[t-1] \cup \widetilde\Phi_i[t].
\end{equation}

\begin{rem}
% \textbf{(Event Uniqueness and Memory Management)}
The timestamp $\tau$ ensures the uniqueness of flags generated by recurrent changes on the same edge. It facilitates garbage collection: since the propagation lifespan of a flag is bounded by the network diameter, nodes can safely purge records from $\mathcal{H}_i$ that are older than a safety margin (e.g., $n$), thereby preventing unbounded memory growth.
\end{rem}

\begin{figure}[htbp]
    \centering  
    \includegraphics[width=1\linewidth]{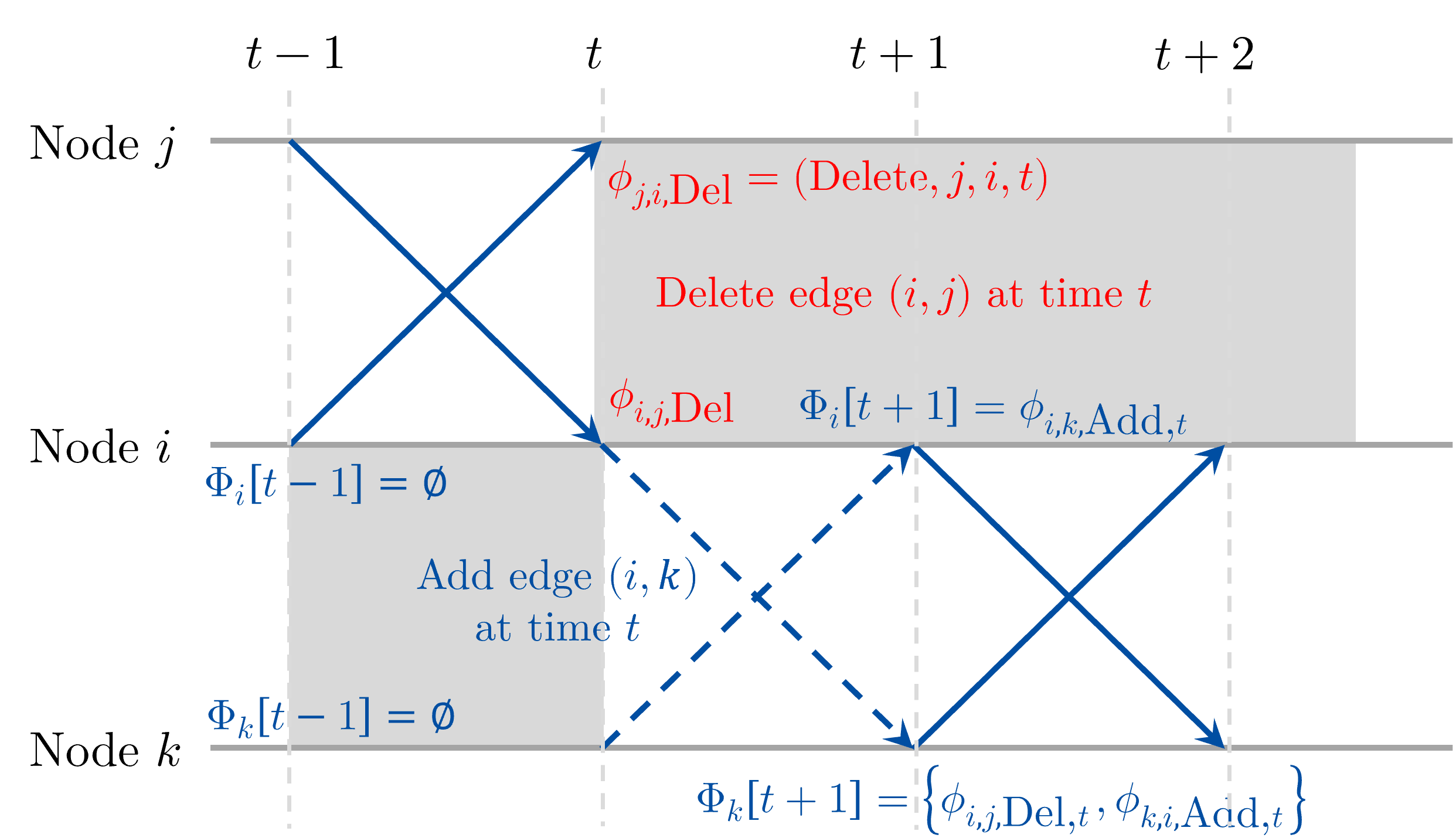}
    \caption{Message flow diagram characterizing the network response to the deletion of edge $(i,j)$ at time $t$ and the addition of edge $(i,k)$ at time $t+1$.}
    \label{Fig:time-line}
    % \vspace{-4mm}
\end{figure}

Fig.~\ref{Fig:time-line} illustrates the generation of change flags when edge $(i,j)$ is deleted at time $t$ and edge $(i,k)$ is added at time $t+1$. At time $t$, node $i$ and $j$ generate the deletion flag $\phi_{i,j,\text{Del},t}$ and propagated to neighbors. At $t+1$, the addition triggers an immediate state exchange (depicted by blue dashed arrows) and the generation of the addition flag $\phi_{i,k,\text{Add},t}$. Consequently, node $i$ aggregates the received deletion flag $\phi_{j,i,\text{Del},t}$ and the locally generated addition flag $\phi_{i,k,\text{Add},t}$ into the active flag set $\Phi_i[t+1]$.

\subsection{Consensus Update Strategy}
\label{sec:consensus_strategy}
Based on the active flags $\Phi_i[t]$, nodes execute a consensus protocol to compute an \emph{intermediate} reachability state, denoted by $\widehat{x}_i[t]$. This state represents the estimate of reachability derived from neighbors' states before accounting for instantaneous local topology changes.

The update follows one of three cases:

\textbf{Case 1: (New Flags Arrived: $\Phi_i[t] \neq \emptyset$)}
Node $i$ integrates information from both historical events and the newly received flags. This update process consists of three steps.

First, it computes a historical reachability state $\overline{x}_i[t]$ based on historical information. Node $i$ computes an intermediate state $\overline{x}_i[t]$ by performing an element-wise logical disjunction (Boolean OR) with these neighbors:
\begin{equation}\label{eq:Historical-Update}
    \overline{x}_i[t] = \bigvee_{j \in \mathcal{N}_i^{hist} \cup \{i\}} x_j[t-1].
\end{equation}
where $\mathcal{N}_i^{hist} = \{j \in \mathcal{N}_i[t-1] \mid \Phi_j[t-1] \cap \mathcal{H}_i[t-1] \neq \emptyset\}$ denotes the subset of neighbors propagating flags that are already present in node $i$'s history set $\mathcal{H}_i[t-1]$. 
This operation ensures that node $i$ performs the standard maximum value update along the propagation path of the old flags, thereby incorporating the updated reachability information from neighbors that have already adapted to previous topological changes.

Second, the algorithm processes the newly received flag set $\widetilde\Phi_i[t]$. For each unique new flag $\phi \in \widetilde\Phi_i[t]$,  the node identifies the set of neighbors $\mathcal{S}(\phi)$ that are carrying that flag, i.e., $\mathcal{S}(\phi) = \{j \in \mathcal{N}_i[t-1] \mid \phi \in \Phi_j[t-1]\}$. 
An intra-flag aggregation is performed 
% using the element-wise logical disjunction (Boolean OR) 
across the set of neighbors $\mathcal{S}(\phi)$ transmitting that specific flag. This computes the maximum reachability supported by the sources of the flag $\phi$, denoted as $x_{\phi}$:
\begin{equation}\label{eq:Vee-Intra-Flag}
    x_{\phi} = \bigvee_{j \in \mathcal{S}(\phi)} x_j[t-1].
\end{equation}
To account for concurrent topology changes, an inter-flag aggregation intersects reachability state of all events reported by the new flags $\phi\in \widetilde\Phi_i[t]$ using the element-wise logical conjunction (Boolean AND). The constraint reachability vector is updated as
\begin{equation}\label{eq:Wedge-Inter-Flag}
    \widetilde{x}_i[t] = \bigwedge_{\phi \in \widetilde\Phi_{i}[t]} x_{\phi},
\end{equation}
where $\bigwedge$ denotes the element-wise logical conjunction.

\begin{lem}
%[Necessary and Sufficient Condition for Reachability Preservation]
\label{lemma:conservative_aggregation}
Under the update rules defined by \eqref{eq:Vee-Intra-Flag} and \eqref{eq:Wedge-Inter-Flag}, node $i$ retains reachability to destination $k$ (indicated by $\widetilde{x}_{ik}[t]=1$) if and only if, for every change flag $\phi$ detected, there exists at least one neighbor $j$ transmitting this flag maintains reachability to $k$ (i.e., $x_{jk}[t-1]=1$).
%$\widetilde{x}_{ik}$ evaluates to 1 if and only if a valid path exists via some neighbor for every concurrent change $\phi$.
\end{lem}
\begin{pf}
    See Appendix \ref{appendix:lemma:conservative_aggregation} for the derivation.
\end{pf} 

Finally, the intermediate reachability state $\widehat{x}_i[t]$ is the intersection of the historical reachability state from the first step and the constraint vector from the second step:
\begin{equation}\label{eq:Final-Consensus}
    \widehat{x}_i[t] = \overline{x}_i[t] \wedge \widetilde{x}_i[t].
\end{equation}

This logic ensures that the impact of new topological changes (encoded in $\widetilde{x}_i[t]$) overrides the historical state $\overline{x}_i[t]$. Specifically, if $\widetilde{x}_{ik}[t]$ becomes zero due to a new edge change, $\widehat{x}_{ik}[t]$ transitions to zero regardless of the historical value. Conversely, if the new change does not affect the shortest path to a destination, the reachability established by the historical update is preserved. 

% The distance vector $\widehat{d}_i[t]$ is subsequently coupled to this reachability update $\widehat{x}_i[t]$, calculated by \eqref{eq:distance-update-general}.

\begin{exmp}
    Consider node $i$ with neighbors $\mathcal{N}_i[t-1]=\mathcal{N}_i[t]=\{j, k, l\}$ and a destination $m$, where the reachability states at time $t-1$ are $x_{im}[t-1]=0$, $x_{jm}[t-1]=1$, $x_{km}[t-1]=0$, and $x_{lm}[t-1]=1$.
    Node $i$ retains a historical flag $\phi_A$ in $\mathcal{H}_i[t-1]$.
    At time $t$, it receives active flag sets from its neighbors:
    $\Phi_j[t-1]=\{\phi_A, \phi_B\}$, $\Phi_k[t-1]=\{\phi_B\}$, and $\Phi_l[t-1]=\{\phi_C\}$.

    The update process at node $i$ proceeds as follows:
    Node $i$ identifies neighbor $j$ as a historical source $(j \in \mathcal{N}_i^{hist})$ because $\Phi_j[t-1] \cap \mathcal{H}_i[t-1] = \{\phi_A\} \neq \emptyset$. 
    % Neighbors $k$ and $l$ only carry new flags ($\phi_B, \phi_C \notin \mathcal{H}_i$).
    Consequently, the intermediate historical state is computed via \eqref{eq:Historical-Update} as
    $$
        \overline{x}_{im}[t] = x_{im}[t-1] \vee x_{jm}[t-1] = 0 \vee 1 = 1.
    $$

    Second, node $i$ identifies the newly received flag set $\widetilde\Phi_i[t] = \{\phi_B, \phi_C\}$.
    For $\phi_B$, which is transmitted by $j,k$, the intra-flag aggregation via \eqref{eq:Vee-Intra-Flag} yields $x_{\phi_B} = x_{jm}[t-1] \vee x_{km}[t-1] = 1 \vee 0 = 1$. For $\phi_C$, transmitted solely by $l$, the aggregation yields $x_{\phi_C} = x_{lm}[t-1] = 1$. These individual flag states are then intersected via \eqref{eq:Wedge-Inter-Flag}:
    $$
        \widetilde{x}_{im}[t] = x_{\phi_B} \wedge x_{\phi_C} = 1 \wedge 1 = 1.
    $$

    Finally, the updated reachability state is determined by intersecting the historical state with the new constraints via \eqref{eq:Final-Consensus}, resulting in
    $$
        \widehat{x}_{im}[t] = \overline{x}_{im}[t] \wedge \widetilde{x}_{im}[t] = 1 \wedge 1 = 1.
    $$
    In this scenario, despite neighbor $k$ reporting unreachability under flag $\phi_B$, the alternative path through $j$ (also carrying $\phi_B$) ensures that reachability to $m$ is maintained.
\end{exmp}

\textbf{Case 2: (No New Flags, Old Flag Held: $\Phi_i[t] = \emptyset \wedge \Phi_i[t-1] \neq \emptyset$)}
If no new flags arrive and node $i$ still holds unpropagated flags from a previous step ($\Phi_i[t-1] \neq \emptyset$), its state must remain constant for one more iteration 
\begin{equation*}
    \widehat{x}_i[t] \leftarrow x_i[t-1].
\end{equation*}

\textbf{Case 3: (No New Flags, No Old Flag Held: $\Phi_i[t] = \emptyset \wedge \Phi_i[t-1] = \emptyset$)}
If no new flags arrive and the node has no local active flags, it performs a standard max-consensus protocol \eqref{eq:standard-max-consensus} to calculate the intermediate reachability state $\widehat{x}_i[t]$.
This allows the algorithm to re-establish connectivity and shortest paths for nodes not directly impacted by the change flags.

Following the reachability calculation, an intermediate distance state $\widehat{d}_i[t]$ is synchronized using a general distance update protocol:
\begin{equation}\label{eq:distance-update-general}
   \widehat{d}_{ik}[t] = \begin{cases} 
    d_{ik}[t-1], & \text{if } x_{ik}[t] = x_{ik}[t-1], \\[4pt]
    \infty, & \begin{aligned}[t]
        &\text{if } x_{ik}[t]=0  \\
        & \wedge x_{ik}[t-1]=1,
        \end{aligned} \\[8pt]
        1 + \min\limits_{j \in \mathcal{N}_i[t-1]} d_{jk}[t-1], & \begin{aligned}[t]
        &\text{if } x_{ik}[t]=1 \\
        & \wedge x_{ik}[t-1]=0.
        \end{aligned}
    \end{cases}
\end{equation}
This protocol handles three scenarios: persistence of valid paths when the reachability state is unchanged; reset to $\infty$ when reachability is lost ($1 \to 0$); 
and re-calculation of the new shortest path when reachability is established ($0 \to 1$).

\subsection{Change Detection and Local State Correction}
\label{Change Detection and Local State Update}
The intermediate states $(\widehat{x}_i[t], \widehat{d}_i[t])$ are derived based on the topology at $t-1$. However, if topological changes occur at time $t$, these estimates must be corrected.
Let $\Delta\mathcal{E}_i[t]$ denote the set of local topology changes detected by node $i$ at time $t$:
\begin{equation*}
    \Delta\mathcal{E}_i[t] = \{(\text{type},i,j) \mid \text{edge } (i,j) \text{ has changed}\}.
\end{equation*}
If $\Delta\mathcal{E}_i[t] \neq \emptyset$, the node executes a correction procedure.

For each detected change $(\text{type}, i, j) \in \Delta\mathcal{E}_i[t]$, node $i$ evaluates whether the change invalidates the current path estimation. This evaluation relies on the \emph{distance difference metric} $\Delta_i^j[t]$, calculated using the states from the previous time step $t-1$\footnote{Note that for edge addition, $d_j[t-1]$ is obtained via an immediate handshake with the new neighbor; for deletion, $d_j[t-1]$ is retrieved from local memory.}:
\begin{equation}\label{eq:Delta}
    \Delta_i^j[t] = d_i[t-1] - d_j[t-1].
\end{equation}
If either $d_i[t-1] = \infty$ or $d_j[t-1] = \infty$, $ \Delta_i^j[t]$ is set to $\infty$.

Based on this metric, the final reachability state $x_i[t]$ is determined by applying a correction rule to the intermediate state $\widehat{x}_i[t]$:
\begin{equation}\label{eq:immediate change}
    x_{ik}[t]=\left\{\begin{array}{ll}
    \widehat{x}_{ik}[t], & \text{if } \Delta_{ik}^j[t] \leq 0, \\[4pt]
    0, & \text{if } \Delta_{ik}^j[t] > 0.
    \end{array}\right.
\end{equation} 

The distance state $d_i[t]$ is synchronized with the corrected reachability. $d_{ik}[t]$ is reset to $\infty$ if $x_{ik}[t]$ is invalidated to $0$, and retains the intermediate value $\widehat{d}_{ik}[t]$ otherwise.

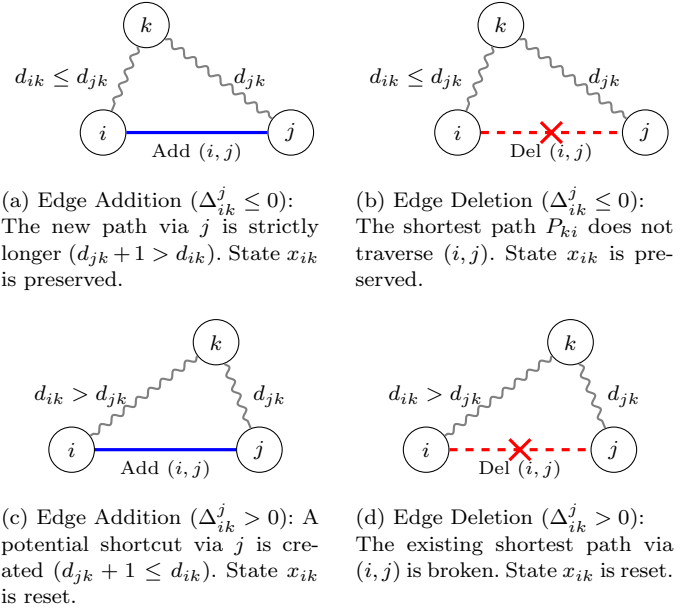
\begin{figure}[htbp]
    \centering
    % 定义样式 (Styles)
    \tikzset{
        % 节点：变大，字体变大
        node_style/.style={circle, draw, minimum size=0.6cm, inner sep=0pt, font=\small},
        % 路径：波浪线，灰色，变粗
        path_edge/.style={draw=gray, text=black, decorate, decoration={snake, amplitude=.4mm, segment length=2mm}, thick},
        % 添加边：蓝色，加粗
        add_edge/.style={draw, very thick, blue, -},
        % 删除边：红色，虚线，加粗
        del_edge/.style={draw, very thick, red, dashed},
        % 叉号：红色，变大
        cross/.style={cross out, draw=red, minimum size=2mm, very thick}
    }

    % Subfigure A
    \begin{subfigure}[b]{0.23\textwidth}
        \centering
        % scale=0.95 稍微放大一点，适应栏宽
        \begin{tikzpicture}[scale=0.95]
            % 稍微拉宽 j 的坐标到 2.6，给中间长文字留空间
            \node[node_style] (i) at (0,0) {$i$};
            \node[node_style] (j) at (2.6,0) {$j$};
            \node[node_style] (k) at (0.6, 1.5) {$k$}; % k 稍微调高一点点保持比例
            
            % 路径标注：使用 midway, above left/right 自动对齐
            \draw[path_edge] (k) -- node[midway, left, font=\footnotesize, xshift=-2pt] {$d_{ik}\leq d_{jk}$} (i);
            \draw[path_edge] (k) -- node[midway, right, font=\footnotesize, xshift=2pt] {$d_{jk}$} (j);
            
            % 蓝色实线 + 距离标号 1
            \draw[add_edge] (i) -- node[below, font=\scriptsize, color=black] {Add $(i,j)$} (j);
        \end{tikzpicture}
        \caption{Edge Addition ($\Delta_{ik}^j \leq 0$): \\
        The new path via $j$ is strictly longer ($d_{jk}+1 > d_{ik}$). State $x_{ik}$ is preserved.}
        \label{fig:sub_a}
    \end{subfigure}
    \hfill    
    % Subfigure B
    \begin{subfigure}[b]{0.23\textwidth}
        \centering
        \begin{tikzpicture}[scale=0.95]
            \node[node_style] (i) at (0,0) {$i$};
            \node[node_style] (j) at (2.6,0) {$j$};
            \node[node_style] (k) at (0.6, 1.5) {$k$};
            
            \draw[path_edge] (k) -- node[midway, left, font=\footnotesize, xshift=-2pt] {$d_{ik}\leq d_{jk}$} (i);
            \draw[path_edge] (k) -- node[midway, right, font=\footnotesize, xshift=2pt] {$d_{jk}$} (j);
            
            % 红色虚线 + 叉号 (删除边通常不标距离，或者标原来的1也可，这里保持清爽不标)
            \draw[del_edge] (i) -- node[below, font=\scriptsize, color=black] {Del $(i,j)$} (j);
            \node[cross] at ($(i)!0.5!(j)$) {};
        \end{tikzpicture}
        \caption{Edge Deletion ($\Delta_{ik}^j \leq 0$):\\
        The shortest path $P_{ki}$ does not traverse $(i,j)$. State $x_{ik}$ is preserved.}
        \label{fig:sub_b}
    \end{subfigure}

    \vspace{3mm}

    % Subfigure C
    \begin{subfigure}[t]{0.23\textwidth}
        \centering
        \begin{tikzpicture}[scale=0.95]
            \node[node_style] (i) at (0,0) {$i$};
            \node[node_style] (j) at (2.6,0) {$j$};
            \node[node_style] (k) at (2.0, 1.5) {$k$}; % k 移到右侧
            
            % 这里的 k 离 j 近，所以左边长，右边短
            \draw[path_edge] (k) -- node[midway, left, font=\footnotesize, xshift=-2pt] {$d_{ik}>d_{jk}$} (i);
            \draw[path_edge] (k) -- node[midway, right, font=\footnotesize, xshift=2pt] {$d_{jk}$} (j);
            
            % 蓝色实线 + 距离标号 1
            \draw[add_edge] (i) -- node[below, font=\scriptsize, color=black] {Add $(i,j)$} (j);
        \end{tikzpicture}
        \caption{Edge Addition ($\Delta_{ik}^j > 0$): A potential shortcut via $j$ is created ($d_{jk}+1 \leq d_{ik}$). State $x_{ik}$ is reset.}
        \label{fig:sub_c}
    \end{subfigure}
    \hfill
    % Subfigure D
    \begin{subfigure}[t]{0.23\textwidth}
        \centering
        \begin{tikzpicture}[scale=0.95]
            \node[node_style] (i) at (0,0) {$i$};
            \node[node_style] (j) at (2.6,0) {$j$};
            \node[node_style] (k) at (2.0, 1.5) {$k$};
            
            \draw[path_edge] (k) -- node[midway, left, font=\footnotesize, xshift=-2pt] {$d_{ik}>d_{jk}$} (i);
            \draw[path_edge] (k) -- node[midway, right, font=\footnotesize, xshift=2pt] {$d_{jk}$} (j);
            \draw[del_edge] (i) -- node[below, font=\scriptsize, color=black] {Del $(i,j)$} (j);
            \node[cross] at ($(i)!0.5!(j)$) {};
        \end{tikzpicture}
        \caption{Edge Deletion ($\Delta_{ik}^j > 0$):\\
        The existing shortest path via $(i,j)$ is broken. State $x_{ik}$ is reset.}
        \label{fig:sub_d}
    \end{subfigure}

    \caption{Illustration of the local state correction logic. Wavy lines represent multi-hop paths ($d_{ik}, d_{jk}$), distinguishing them from the direct edge $(i,j)$ with unit weight being added (solid blue) or deleted (dashed red).}
    \label{fig:logic_illustration}
\end{figure}

The correction rule \eqref{eq:immediate change} serves as a conservative invalidation mechanism based on the topological relationship between nodes $k$, $i$, and $j$, as illustrated in Fig.~\ref{fig:logic_illustration}.
Intuitively, if $\Delta_{ik}^j[t] \leq 0$, node $k$ is closer to $i$ than to $j$. This implies that the shortest path from $k$ to $i$ does not utilize edge $(i,j)$; thus, its modification does not affect the current state $x_{ik}$ (see Figs. \ref{fig:sub_a} and \ref{fig:sub_b}).
Conversely, if $\Delta_{ik}^j[t] > 0$, edge $(i,j)$ either constitutes the final hop of the shortest path (in deletion) or introduces a potential shortcut (in addition), necessitating a state reset to force re-evaluation (see Figs. \ref{fig:sub_c} and \ref{fig:sub_d}).
A detailed derivation of this rationale is provided in Appendix~\ref{appendix:Design Rationale}.

Simultaneously, a unique change flag $\phi=(\text{type},i,j,t)$ is generated for this change event and added to the locally generated flag set $\Delta\Phi_i[t]$.
%denotes the set of locally generated change flags resulting from these events, initialized as $\Delta\Phi_i[0] = \emptyset$.}
After processing all local changes, the active flag set $\Phi_i[t]$ is updated by determining the union of locally generated flag set $\Delta\Phi_i[t]$ and newly received flag set $\widetilde{\Phi}_i[t]$:
\begin{equation}\label{eq:local_flag_insertion}
\Phi_i[t] = \widetilde{\Phi}_i[t] \cup \Delta\Phi_i[t].
\end{equation}

If no local changes are detected ($\Delta\mathcal{E}_i[t] = \emptyset$), the intermediate states are adopted as final, i.e., $x_i[t] = \widehat{x}_i[t]$ and $d_i[t] = \widehat{d}_i[t]$. 

The complete execution procedure for calculating the updated states and generating the set of local flags is presented in Algorithm~\ref{alg:node_local_reaction}.

\begin{algorithm}[h!]
\caption{Local State Update on Edge Change for Node $i$ at Time~$t$}
\label{alg:node_local_reaction}
\begin{algorithmic}[1]
\Procedure{OnEdgeChange}{$\Delta\mathcal{E}_i[t], \widehat{x}_i[t], \widehat{d}_i[t]$}
    %\State \textbf{Input:} 
    %Previous states $x_i[t-1], d_i[t-1]$; 
    %Set of local changes $\Delta\mathcal{E}_i[t]$;
    %\State \textbf{Output:} 
    %Updated states $x_i[t], d_i[t]$; Locally generated flag set~$\Delta \Phi_i$.
    
    \State $\Delta \Phi_i[t] \gets \emptyset$

    \State \textbf{/* Process all local changes sequentially */}
    \ForAll{$(\text{type},i,j) \in \Delta\mathcal{E}_i[t]$}
        
        \If{$\text{type} == \text{ADD}$}
            \State Receive $d_j[t-1]$ from new neighbor $j$ \Comment{Immediate handshake}
        \Else \Comment{$\text{DELETE}$}
            \State Retrieve $d_j[t-1]$ from local memory
        \EndIf

        \State Compute $\Delta_i^j[t]$ via \eqref{eq:Delta}
        \State Update $x_i[t]$ via \eqref{eq:immediate change}

        \State $\phi \gets (\text{type}, i, j,t)$ \Comment{Create unique event flag}
        \State $\Delta \Phi_i[t] \gets \Delta \Phi_i[t] \cup \{\phi\}$
    \EndFor
    % % 在 Algorithm 1 中替换原来的红色部分
    % \State $d_{ik}[t] \gets \begin{cases} \infty, & \text{if } x_{ik}[t]=0 \\ \widehat{d}_{ik}[t], & \text{otherwise} \end{cases}$
    % 在 Algorithm 1 中替换原来的红色部分
    \State Set $d_{ik}[t] \gets \infty$ if $x_{ik}[t]=0$, else $d_{ik}[t] \gets \widehat{d}_{ik}[t]$
    % \State \red{Update $d_i[t]$ via \eqref{eq:distance-update-general}}

    \State \Return $x_i[t], d_i[t], \Delta \Phi_i[t]$
\EndProcedure
\end{algorithmic}
\end{algorithm}

\subsection{Algorithm and Analysis}
The complete distributed protocol integrates the flag propagation mechanism (Sec.~\ref{sec:propagation}), the consensus update strategy (Sec.~\ref{sec:consensus_strategy}), and the local change detection (Sec.~\ref{Change Detection and Local State Update}). 
Algorithm~\ref{alg:main_loop} outlines the execution flow at a generic node $i$ during time step $t$.
In each iteration, nodes first exchange information with neighbors to update their flag sets, reachability states and distance states, and subsequently handle any locally detected topological events.
Following the algorithmic description, we provide the theoretical convergence analysis.

\begin{algorithm}[h!]\caption{Distributed Reachability and Distance Update for Node $i$ at Time $t$}\label{alg:main_loop}
\begin{algorithmic}[1]
    \Statex \textbf{Input:} Previous states $x_j[t-1], d_j[t-1], \Phi_j[t-1]$ for all $j \in \overline{\mathcal{N}}_i[t-1]$;
    State history $\mathcal{H}_i[t-1]$;
    Set of local changes $\Delta\mathcal{E}_i[t]$
    \Statex \textbf{Output:} Updated states $x_i[t],d_i[t]$
    % \Repeat
        % \State $t \gets t + 1$
        % \State $\text{converged} \gets \text{true}$
            % \Statex \textbf{/* Store previous state for comparison */}
            % \State $x_i[t] \gets x_i[t-1]$
            % \State $d_i[t] \gets d_i[t-1]$    
            \Statex \textbf{/* Change Flag Propagation */}
            % \State Calculate neighbor flag union $\Psi_i[t]$ via \eqref{eq:flag_union}
            \State Calculate newly received flag set $\widetilde{\Phi}_i[t]$ via \eqref{eq:phi_tilde_update}
            \State Initialize active flag set $\Phi_i[t] \gets \widetilde{\Phi}_i[t]$ 
            % \Comment{Initialize with propagated flags}
    
            \Statex \textbf{/* Consensus Update Strategy */}
            \If{$\Phi_i[t] \neq \emptyset$} \Comment{Case 1: New flags arrived}
                % \State $\text{converged} \gets \text{false}$
                % \State $x_k[t] \gets x_k[t-1]$ \Comment{Start with previous state}
            % \For{each unique flag $\phi \in \Phi_i[t]$}
            %     \State $x_{\phi} \gets \mathbf{0}$ \Comment{Initialize state for max aggregation}
            %     \For{$j \in \{j' \in \mathcal{N}_i \mid \phi \in \Phi_{j'}[t-1]\}$}
            %         \State $x_{\phi} \gets \max(x_{\phi}, x_j[t-1])$ \Comment{Element-wise max}
            %     \EndFor
            %     \State $\widetilde{x}_i = x_i[t-1] \wedge \left( \bigwedge_{\phi \in \Phi_{i}[t]} x_{\phi} \right)$
            % \EndFor
            \State Calculate intermediate state $\overline{x}_i$ via \eqref{eq:Historical-Update} 
            % for all $\mathcal{N}_i^{hist}$ carrying historical information
            \State Calculate intra-flag aggregation $x_{\phi}$ via \eqref{eq:Vee-Intra-Flag} 
            % for all $\phi \in \widetilde\Phi_i[t]$
            \State Calculate inter-flag aggregation $\widetilde{x}_i$ via \eqref{eq:Wedge-Inter-Flag}
            % \State Update $x_i[t]$ by an intersection step \eqref{eq:Final-Consensus}
            \State $\widehat{x}_i[t] \gets \overline{x}_i \wedge \widetilde{x}_i$ \Comment{Intermediate reachability}
        \ElsIf{$\Phi_i[t-1] \neq \emptyset$} \Comment{Case 2: Old flag held}
            \State $\widehat{x}_i[t] \gets x_i[t-1]$ \Comment{Local flag persistence}
        \Else \Comment{Case 3: No new flags, no old flag held}
            \State $\widehat{x}_i[t] = \bigvee_{j \in \overline{\mathcal{N}}_i[t]} x_j[t-1]$ 
            % \Comment{\orange{Max-consensus}}
            % \State $x_i[t] \gets \mathbf{0}$ 
            % \For{$j \in \mathcal{N}_i$}
            %     \State $x_i[t] \gets \max(x_i[t], x_j[t-1])$
            % \EndFor
        \EndIf
    
        \Statex \textbf{/* Distance State Update */}
        \State Update $\widehat{d}_{i}[t]$ based on $\widehat{x}_i[t]$ via \eqref{eq:distance-update-general}
        % \If{$x_i[t] \neq x_i[t-1]$}
        %     \State $\text{converged} \gets \text{false}$
        % \EndIf
        
        \Statex \textbf{/* Local State Correction on Edge Change */}
        \If{$\Delta\mathcal{E}_i[t] \neq \emptyset$} \Comment{Change events detected}
            \State $[x_i[t], d_i[t], \Delta\Phi_i[t]] \gets$
            \State $\quad \quad \Call{OnEdgeChange}{\Delta\mathcal{E}_i[t], \widehat{x}_i[t], \widehat{d}_i[t]}$
            % \State Update active flag set $\Phi_i[t]$ via \eqref{eq:local_flag_insertion} 
            \State $\Phi_i[t] \gets \widetilde{\Phi}_i[t] \cup \Delta\Phi_i[t]$ \Comment{Merge locally generated flags}
        \Else
            \State $x_i[t] \gets \widehat{x}_i[t]$ \Comment{No local change, keep intermediate}
            \State $d_i[t] \gets \widehat{d}_i[t]$
        \EndIf
        \State Update history $\mathcal{H}_i[t]$ via \eqref{eq:history_update}
        \State Send out $x_i[t], d_i[t], \Phi_i[t]$ to all neighbors $\mathcal{N}_i[t]$
        
% \Until{$\text{converged}$}

\end{algorithmic}\end{algorithm}

% \begin{thm}\label{thm:convergence}
%     %[Robustness and Eventual Convergence]
%     Let the network topology $\mathcal{G}[t] = (\mathcal{V}, \mathcal{E}[t])$ be subject to an arbitrary sequence of edge addition and deletion events, where each event $c$ occurring at time $T_c$ generates a flag $\phi_c$. If there exists a time $T_{final}$ after which the graph topology stabilizes (i.e., $\forall t \ge T_{final}$, $\mathcal{G}[t] = \mathcal{G}[T_{final}]$), the algorithm is guaranteed to converge to the correct states $(x^*, d^*)$ corresponding to the final graph $\mathcal{G}[T_{final}]$ within a finite time.
% \end{thm} 
\begin{thm}[Dynamic Re-convergence]\label{thm:convergence}
Consider a network $\mathcal{G}[t]$ undergoing an arbitrary sequence of topological changes. 
If the topology remains unchanged for an interval of $2n$ communication rounds starting from time $T$, the algorithm guarantees convergence to the correct values $(x^*[T], d^*[T])$ of $\mathcal{G}[T]$ within at most $2n$ communication rounds.
% , regardless of the any arbitrary state $(x[T], d[T])$ at the beginning of this interval.
\end{thm}
\begin{pf}
    The complete proof is presented in Appendix \ref{appendix:thm:convergence}.
\end{pf}
% \begin{thm}[Dynamic Re-convergence]\label{thm:convergence}
% For any arbitrary state $(x[T_{stab}], d[T_{stab}])$, the algorithm guarantees convergence to the correct values $(x^*, d^*)$ of $\mathcal{G}_{final}$ within at most $2n$ communication rounds.
% \end{thm}

\begin{rem}
% Theorem~\ref{thm:convergence} establishes the property of \emph{self-stabilization}. 
Unlike static algorithms that require specific initialization (e.g., $d_i=\infty$ for all nodes), our algorithm recovers the correct states from any arbitrary state configuration resulting from prior events, without requiring global re-initialization.
\end{rem}

\subsection{Privacy Preservation}
\begin{figure}[htbp]
    \centering
    \begin{minipage}{0.45\linewidth}
        \includegraphics[width=\linewidth]{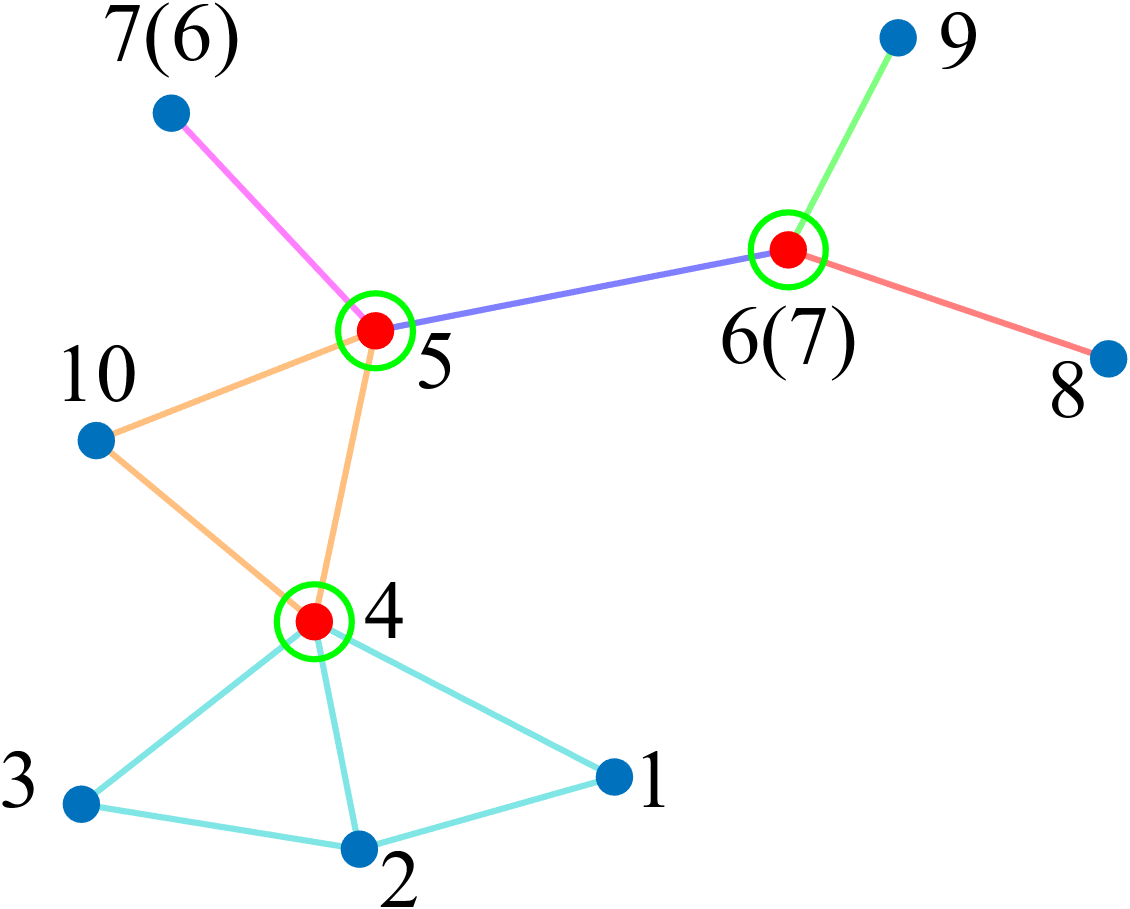}
    \end{minipage}
    \hfill
    \begin{minipage}{0.45\linewidth}
    \includegraphics[width=\linewidth]{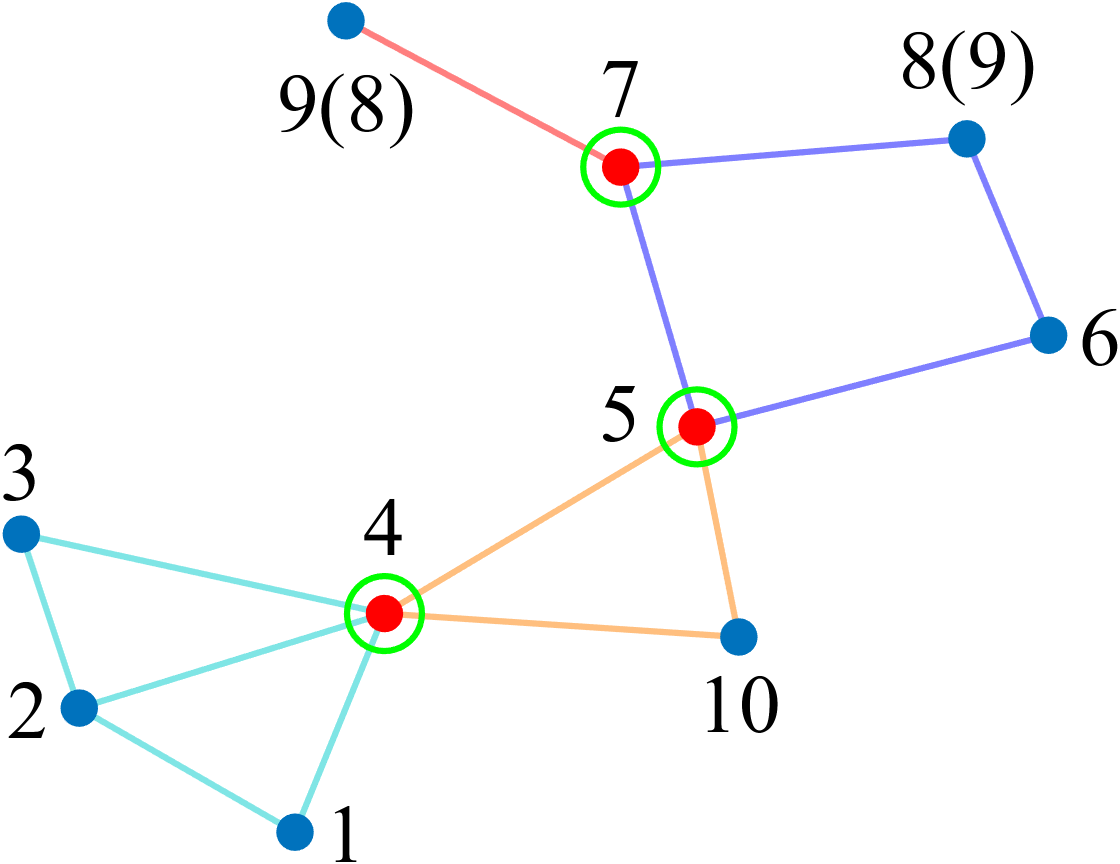}
    \end{minipage}
    \caption{Compared to Fig.~\ref{Fig:easy}, the two topologies differ in the addition of edge $(5,10)$, the removal of edge $(6,7)$, and other edge variations, yet both maintain the same distance state at node $2$. The notation $i(j)$ indicates that $i$ and $j$ are interchangeable nodes.}
    \label{pricacy}
\end{figure}

The incremental update mechanism inherits the privacy-preserving properties of the max-consensus protocol. Nodes only exchange state vectors $x_i, d_i$ and flags $\phi$, never their adjacency lists $\mathcal{N}_i$. As demonstrated in Fig.~\ref{pricacy}, this information is insufficient for a node to reconstruct the topology outside their 2-hop range.

Take node~$2$ as an example in Fig.~\ref{Fig:easy}. The distance state $d_2$ converges to
$$d_2=[1,0,1,1,2,3,3,4,4,2].$$ 
Node $2$ receives the converged distance states from its neighbors $\mathcal{N}_2=\{1,3,4\}$, i.e.,
$$\begin{array}{cc}
d_1=d_3=[1,0,1,1,2,3,3,4,4,2],\\ [4pt]
d_4=[1,1,1,0,1,2,2,3,3,1].
\end{array}$$

Based solely on the information at node $2$ after convergence, i.e., $d_1, d_2, d_3, d_4$, the topology beyond the 2-hop neighbors cannot be determined. Specifically, the connections between node $5$ and $10$, as well as the connections among nodes $5, 6, 7, 8, 9$ remain unknown. As demonstrated in Fig.~\ref{pricacy}, two distinct topologies can result in identical distance state at node $2$.

\section{Distributed AP Identification in Time-Varying Networks}
\label{Section:Distributed AP Identification}
This section details the procedure for a node to determine its AP status. Unlike static algorithms that execute a single computation pass, our dynamic framework triggers AP identification when the node reaches a stable local state.

\subsection{Trigger Condition}
Node $i$ starts the identification algorithm at time $t$ if and only if two stability conditions are met:
\begin{enumerate}
    \item The node holds no change flags ($\Phi_i[t] = \emptyset$) and receives no flags from any neighbor ($\Phi_j[t-1] = \emptyset, \forall j \in \mathcal{N}_i[t-1]$).
    \item Its local reachability and distance vectors remain unchanged from the previous step ($x_i[t] = x_i[t-1]$ and $d_i[t] = d_i[t-1]$).
\end{enumerate}
If a new flag arrives subsequently, the node resumes the incremental update process via Algorithm~\ref{alg:main_loop} and re-evaluates its status once stability is regained.
Given that the identification process operates under these stable conditions, the time index $t$ is omitted in the subsequent description for brevity.

\subsection{Identification Criterion and Biconnectivity Check}
Once the local state stabilizes, node $i$ applies the distributed criterion established in our preliminary work \citep{Xie2025CDC}. This method utilizes the local distance states to infer the connectivity of the node's neighborhood.
% we adopt the identification framework established in our preliminary work \citep{Xie2025CDC}, which utilizes the local distance information to infer topological features.

We define $\mathcal{P}_i$ as the set of neighbor pairs $\{j,k\}$ that satisfy the sufficient conditions for belonging to a common cycle containing $i, j$, and $k$. As derived in \cite{Xie2025CDC}, this set is constructed locally using the distance difference metric $\Delta$ (Eq.~\eqref{eq:Delta}):
\begin{equation}\label{setP}
    \mathcal{P}_i = \left\{ \{j,k\} \in \binom{\mathcal{N}_i}{2} \;\Bigg|\; 
    \begin{aligned}
    &\Delta_{ik}^{j}=\Delta_{ij}^{k}=0 \text{, or } \\
    &\exists p\in\mathcal{V}\setminus \overline{\mathcal{N}}_i, \Delta_{ip}^{j}\geq 0 \wedge \Delta_{ip}^{k}\geq 0
    \end{aligned}
    \right\}.
\end{equation}
Based on this set, the node determines its AP status via the following theorem.

\begin{thm} [\citealp{Xie2025CDC}]
\label{theorem: Distributed identification}
A node $i$ is not an AP if and only if there exists a set $\mathcal{Q}_i$ such that:
\begin{enumerate}
\item $\mathcal{Q}_i \subseteq \mathcal{P}_i$;
\item The Union-Find structure derived from $\mathcal{Q}_i$ connects all neighbors in $\mathcal{N}_i$ into a single component (i.e., $|\mathcal{C}_i|=1$);
\item $\mathcal{Q}_i$ covers all nodes in $\mathcal{N}_i$.
\end{enumerate}
\end{thm}

% This criterion enables node $i$ to verify if its removal would disconnect its neighbors using only the locally maintained distance vector $d_i$ and neighbor IDs.

Beyond local self-identification, this framework naturally extends to monitoring global network biconnectivity. As detailed in \cite{Xie2025CDC}, by employing an auxiliary boolean state variable propagated via the standard max-consensus protocol, nodes can aggregate the identification results. The network is assessed as biconnected if and only if the consensus result indicates that no node has identified itself as an articulation point.

% \subsection{Network-Wide Biconnectivity Monitoring}
% Beyond local self-identification, the proposed framework enables monitoring of the global network biconnectivity.
% As detailed in \cite{Xie2025CDC}, this is achieved by introducing an auxiliary state variable to track the completion and results of the AP identification process. Once node $i$ determines its local status via Theorem~\ref{theorem: Distributed identification}, it propagates a boolean flag using the standard max-consensus protocol. 

% Consequently, every node in the network eventually aggregates the AP status of all other reachable nodes. The network is assessed as biconnected if and only if the consensus result indicates that no node has identified itself as an articulation point. 

\section{Simulation Experiments}
% The experiments are designed to test the algorithm's correctness, efficiency, and robustness by subjecting it to a series of controlled network topological events.

The simulations are based on the 10-node network shown in Fig.~\ref{Fig:easy}. To quantify correctness, a ground truth was established. For any time step $t$, the ground truth $(x^*[t], d^*[t])$ represents the theoretically correct states corresponding to the instantaneous network topology $\mathcal{G}[t]$. The performance metric is the state error, defined as the L1 norm between the algorithm's instantaneous state $x[t]$ and the ground truth state $x^*[t]$. An error of zero signifies convergence to the correct state:
$$\text{Error}[t] = \sum_{i \in \mathcal{V}} \sum_{k \in \mathcal{V}} |x_{ik}[t] - x^*_{ik}[t]|$$
In the simulation, consistent with the distributed protocol, each node independently triggers the AP identification procedure upon detecting local stability, i.e., empty flag sets and constant local states.

The algorithm's performance was first evaluated under elementary topological changes, representing isolated link failure (``bad news'') and establishment (``good news'') events.

\begin{figure}[htbp]
    \centering
    \begin{subfigure}[t]{0.47\linewidth}
        \centering
        \includegraphics[width=\linewidth]{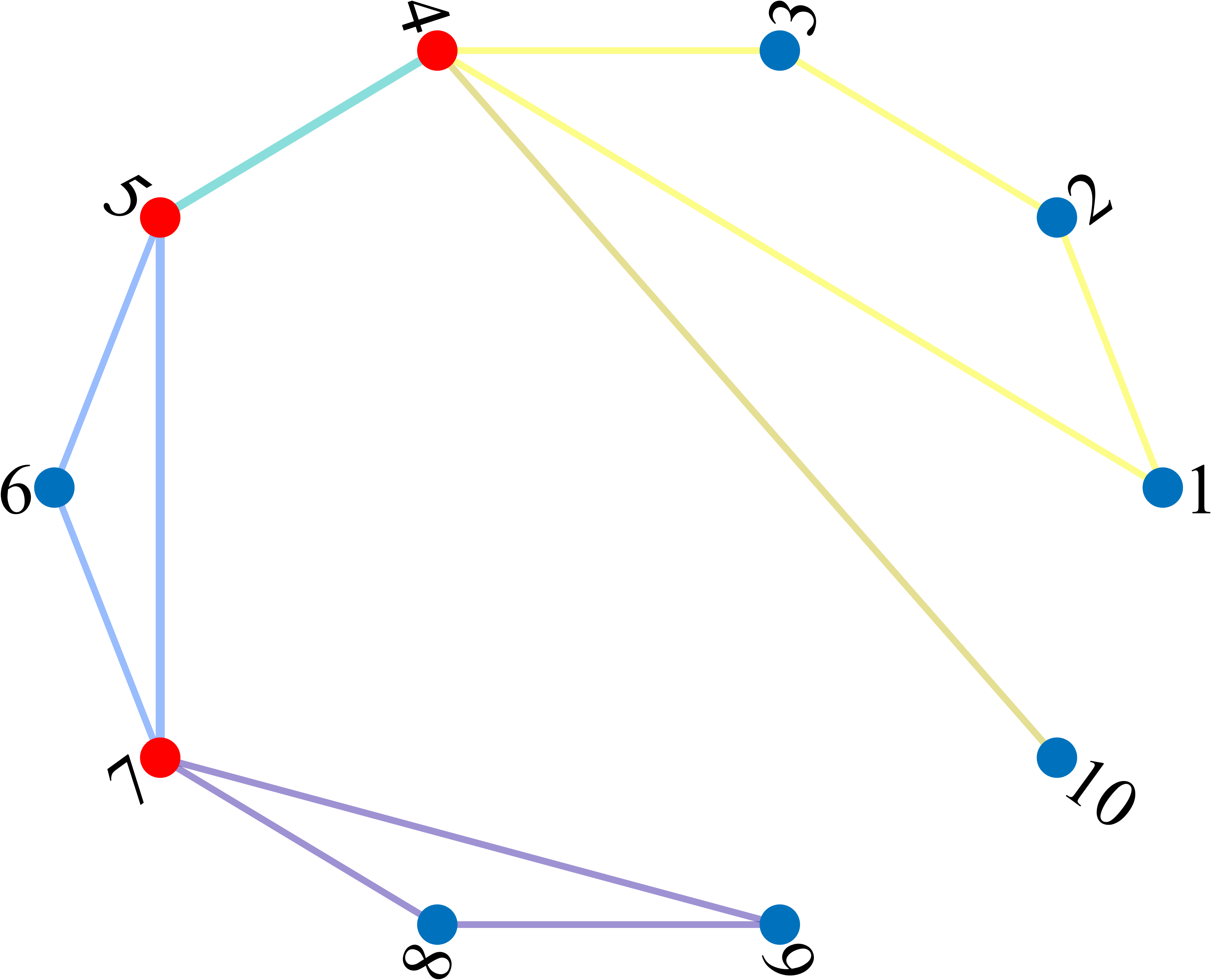}
        \caption{Topology showing deletion of edge $(2, 4)$.}
        \label{fig:del_topo}
    \end{subfigure}
    \hfill
    \begin{subfigure}[t]{0.47\linewidth}
        \centering
        \includegraphics[width=\linewidth]{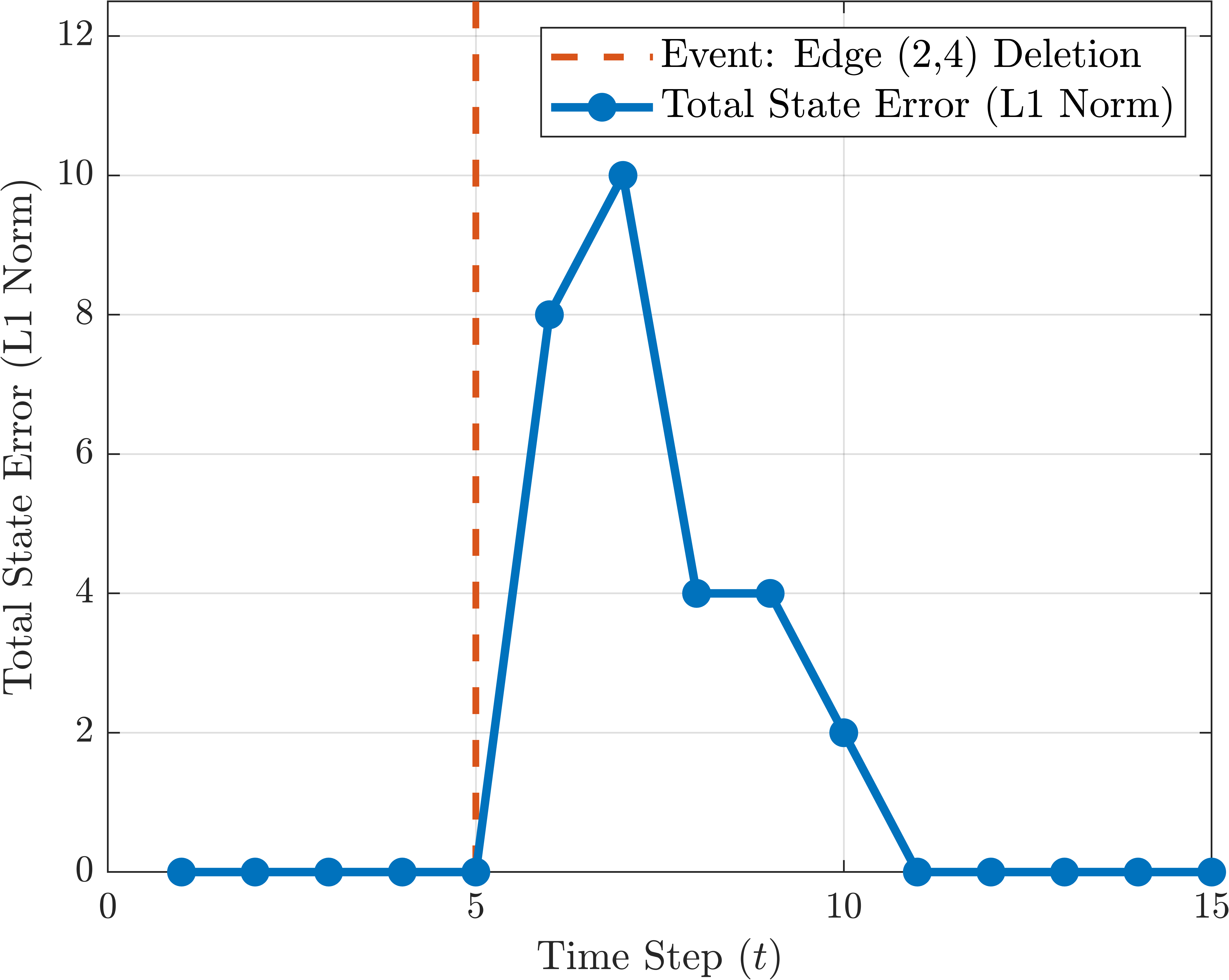}
        \caption{Total state error convergence.}
        \label{fig:del_error}
    \end{subfigure}
    \caption{Experimental results for single edge deletion event at $t=5$.}
    \label{Fig:Edge_Deletion_Combined}
\end{figure}

In a single-edge deletion scenario, the edge $(2, 4)$ was removed at $t=5$, as illustrated in Fig.~\ref{fig:del_topo}. As shown in the convergence plot Fig.~\ref{fig:del_error}, this event induced an immediate state error. The propagation of the deletion flag triggered the Case 1 update, and the error was observed to converge to zero within 6 steps. Once local stability was regained, each node verified that the set of APs remained unchanged as $\{4, 5, 7\}$ with the distributed identification criterion.
% 场景1：单边删除

\begin{figure}[htbp]
    \centering
    \begin{subfigure}[t]{0.47\linewidth}
        \centering
        \includegraphics[width=\linewidth]{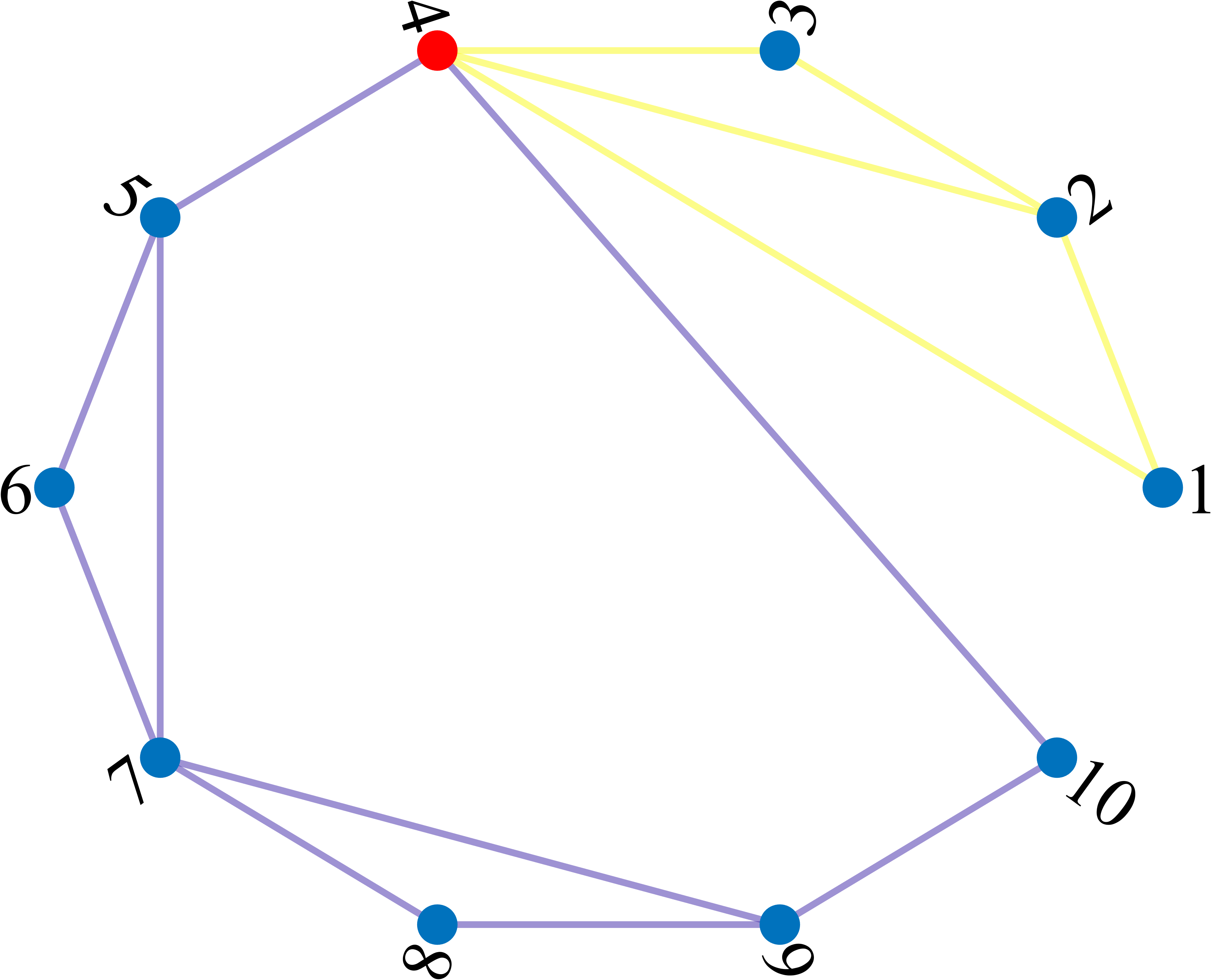}
        \caption{Topology showing addition of edge $(9, 10)$.}
        \label{fig:add_topo}
    \end{subfigure}
    \hfill
    \begin{subfigure}[t]{0.47\linewidth}
        \centering
        \includegraphics[width=\linewidth]{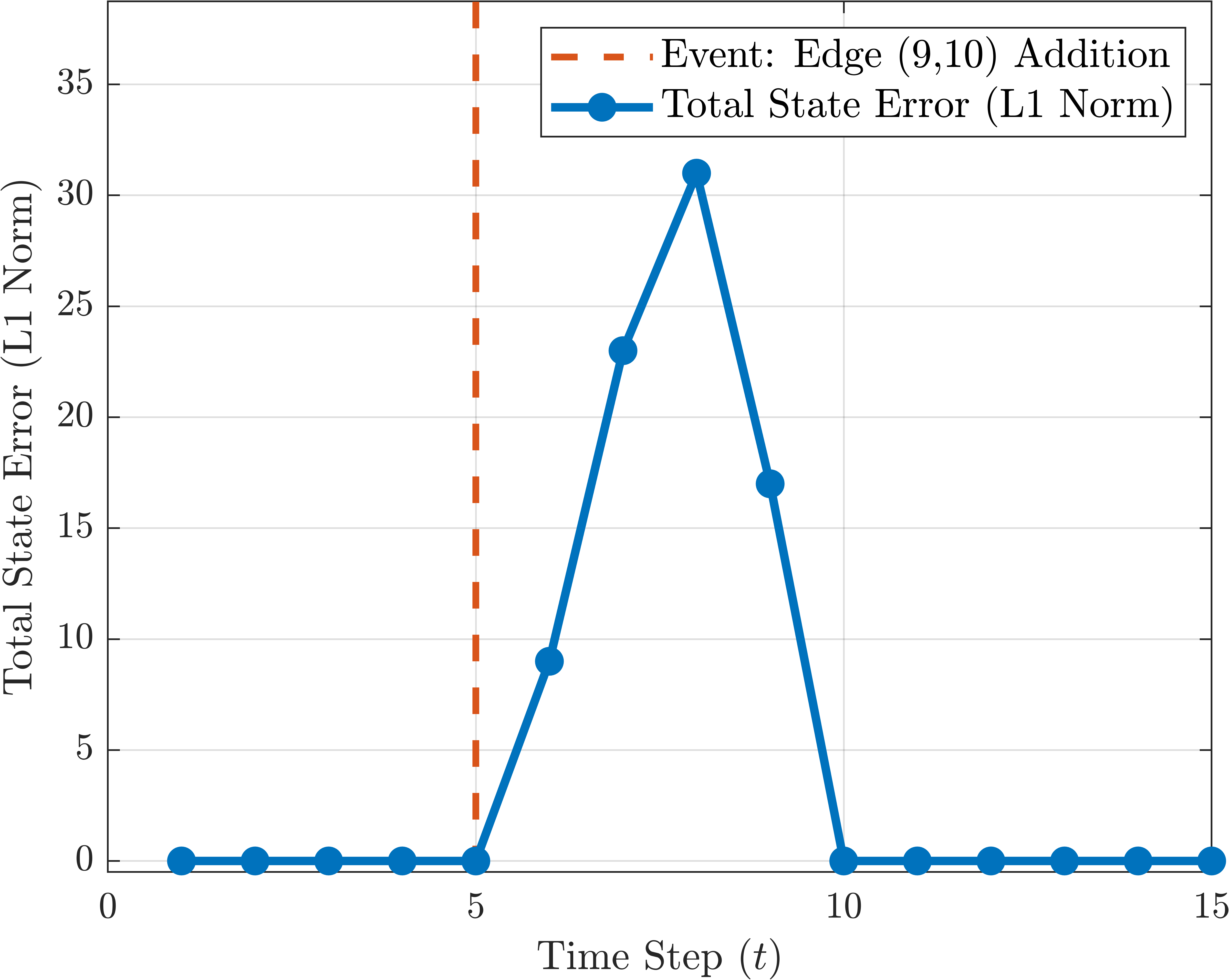}
        \caption{Total state error convergence.}
        \label{fig:add_error}
    \end{subfigure}
    \caption{Experimental results for single edge addition event at $t=5$.}
    \label{Fig:Edge_Addition_Combined}
\end{figure}

A complementary experiment, involving the addition of a ``shortcut'' edge $(9, 10)$, produced a similar result, as shown in Fig.~\ref{fig:add_topo}. The flag generated by this addition triggered updates that incorporated the new path, with the network converging to the new states in 5 steps as shown in Fig.~\ref{fig:add_error}. Once local stability was regained, each node correctly updated the AP set from $\{4, 5, 7\}$ to $\{4\}$, successfully detecting the elimination of APs $5$ and $7$.

% 场景2：单边添加

\begin{figure}[htbp]
    \centering
    \begin{subfigure}[b]{0.47\linewidth}
        \centering
        \includegraphics[width=\linewidth]{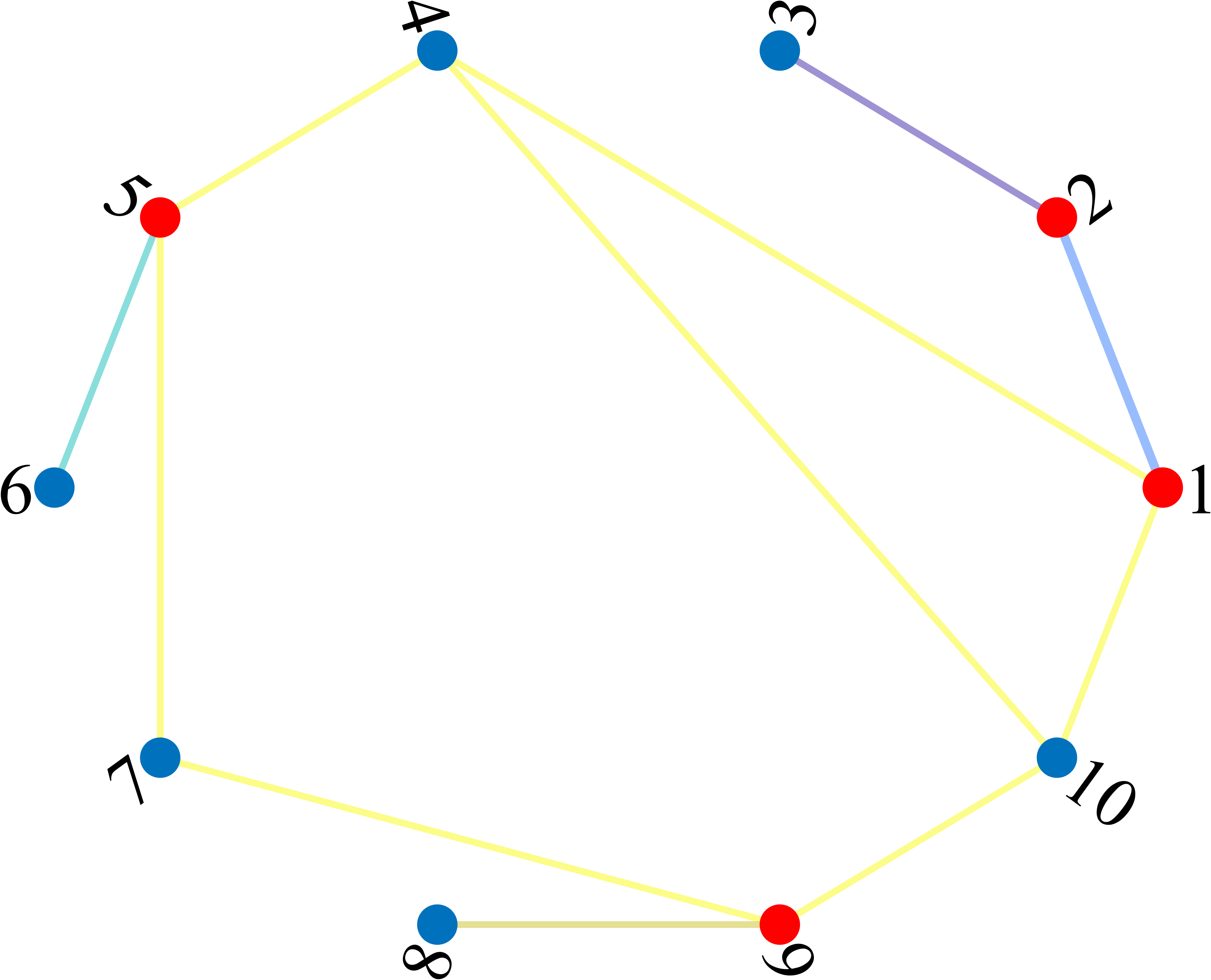}
        \caption{Sequence of topological changes.}
        \label{fig:concurrent_topo}
    \end{subfigure}
    \hfill
    \begin{subfigure}[b]{0.47\linewidth}
        \centering
        \includegraphics[width=\linewidth]{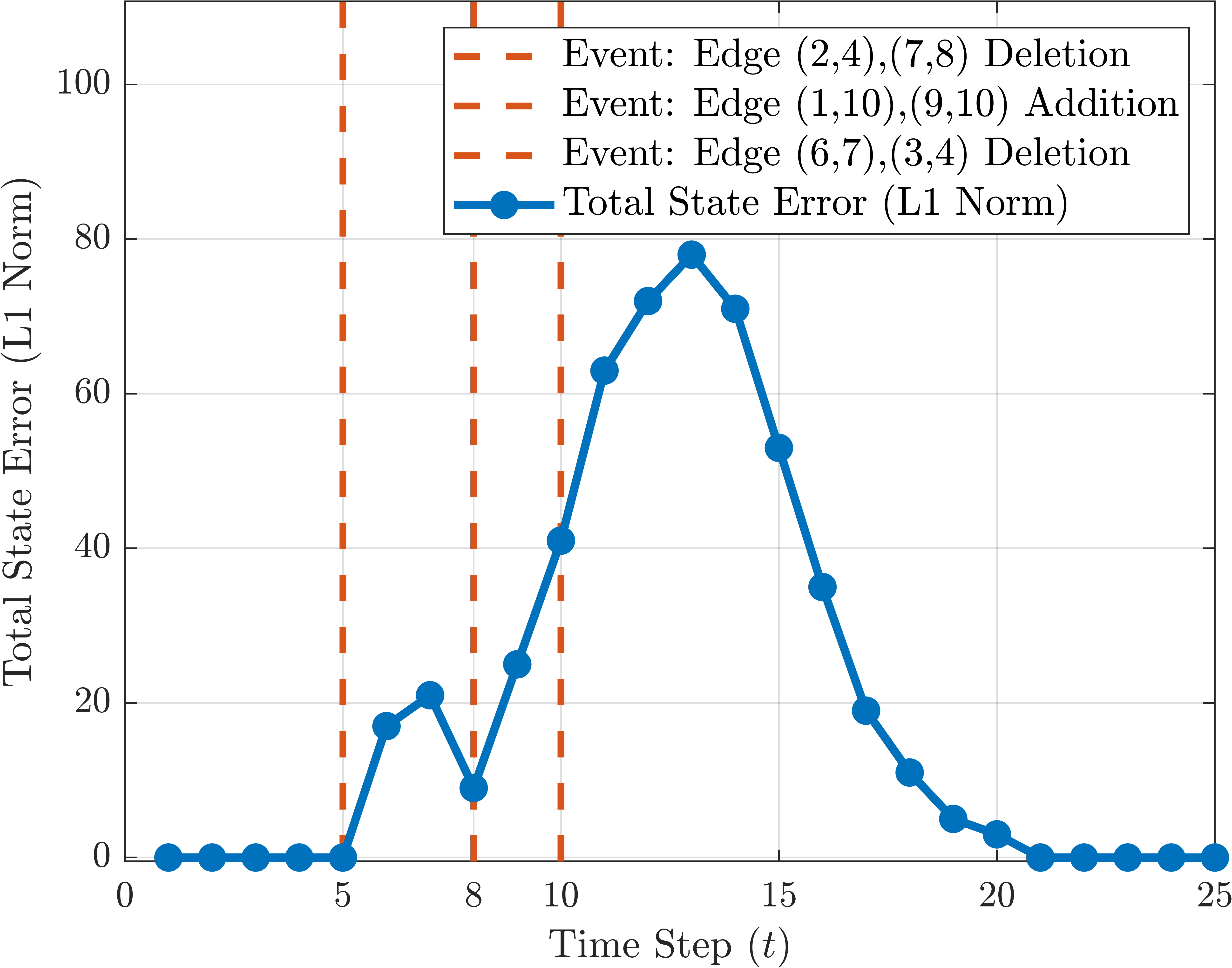}
        \caption{Total state error response under sequential changes.}
        \label{fig:concurrent_error}
    \end{subfigure}
    \caption{Experimental results for concurrent edge addition and deletion events between $t=5$ and $t=10$.}
    \label{Fig:Edge_Continuous_Changes_Combined}
\end{figure}

Further validation involved testing the algorithm's handling of concurrent events. We simulated a sequence of rapid changes summarized in Fig.~\ref{fig:concurrent_topo}: edges $(2, 4)$ and $(7, 8)$ were deleted at $t=5$, edges $(9, 10)$ and $(1,10)$ were added at $t=8$, and edges $(6, 7)$ and $(3, 4)$ were deleted at $t=10$. 
% 场景3：连续并发事件
As depicted in Fig.~\ref{fig:concurrent_error}, the algorithm remained robust and guaranteed convergence to the correct final state at $t=21$.
Each node updated the AP set from $\{4, 5, 7\}$ to $\{1,2,4,9\}$.

\begin{figure}[htbp]
    \centering
    \begin{subfigure}[b]{0.47\linewidth}
        \centering
        \includegraphics[width=\linewidth]{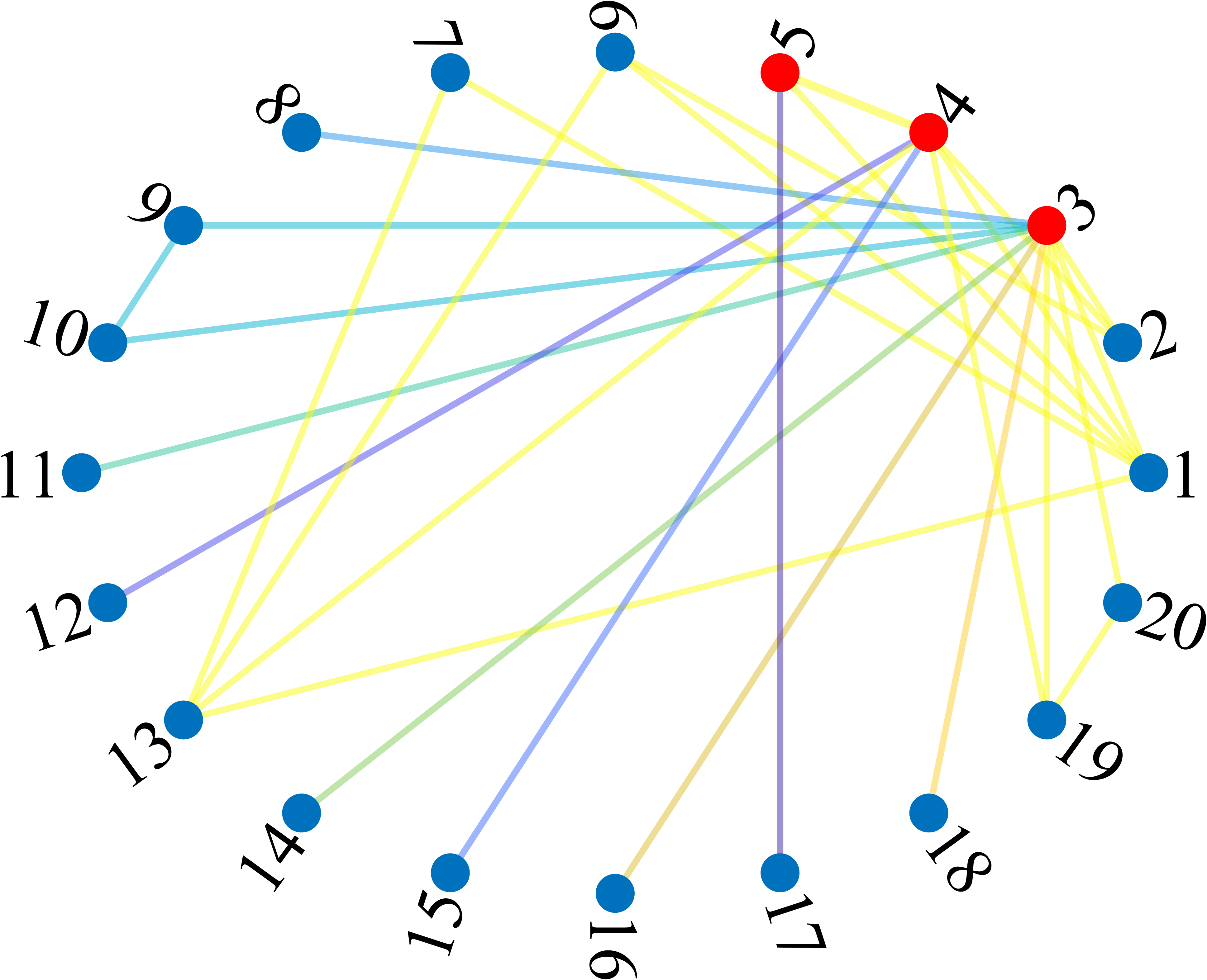}
        \caption{Representative 20-node BA network topology.}
        \label{fig:ba_topo}
    \end{subfigure}
    \hfill
    \begin{subfigure}[b]{0.47\linewidth}
        \centering
        \includegraphics[width=\linewidth]{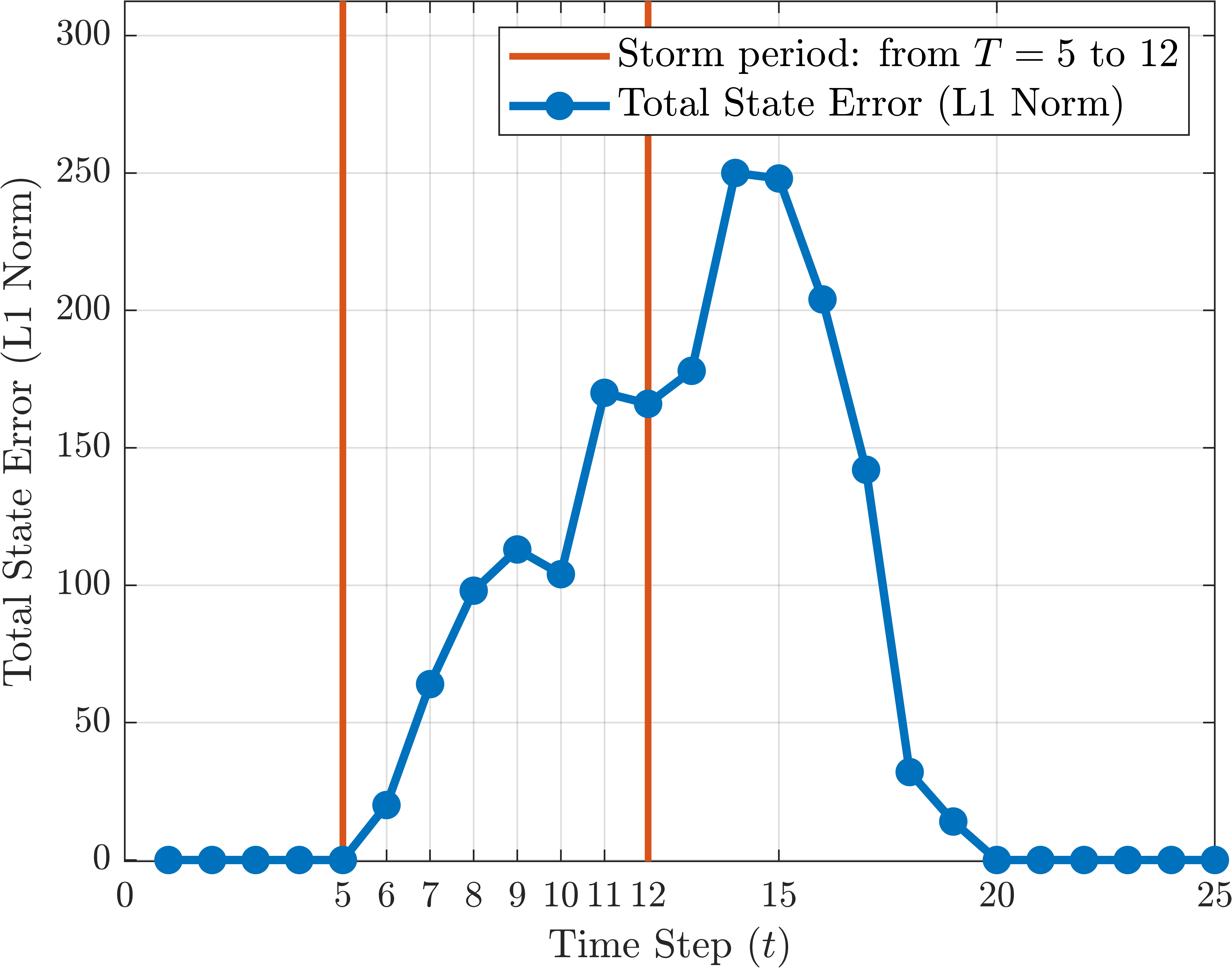}
        \caption{State error response to high-frequency event storm.}
        \label{fig:ba_error}
    \end{subfigure}
    \caption{Experimental results on the 20-node BA network subjected to random changes between $t=5$ and $t=12$.}
    \label{Fig:Edge_Random_Change_Combined}
\end{figure}

To provide a more stressful test of the algorithm's robustness, we simulated an ``event storm'' scenario on a 20-node Barabási-Albert (BA) network (Fig.~\ref{fig:ba_topo}). This test involved high-frequency random changes occurring between $t=5$ to $t=12$. During this window, a random topological change (either an edge addition or deletion) was initiated at each time step. After $t_{final}=12$, the topology was held static.
The results of a single realization are shown in Fig.~\ref{fig:ba_error}. The state error increased during $t \in [5, 12]$ as new state errors were continually introduced by the rapid changes. After the storm stopped at $t_{final}=12$, the algorithm processed all remaining flags and the state error rapidly converged to zero. 
% With the return to a static topology and subsequent local stabilization, the algorithm updates the identified AP set from the initial $\{1, 3, 4\}$ to the final configuration $\{3, 4, 5\}$.

% 场景4：随机风暴
\section{Conclusion}
This paper presents a fully distributed framework for identifying articulation points in time-varying undirected networks. By introducing an  incremental update mechanism to the max-consensus protocol, our algorithm efficiently adapts to topological changes without requiring global re-initialization. Theoretical analysis confirms that the algorithm guarantees eventual convergence to the correct network state provided the topology remains static for a finite interval. Furthermore, the proposed approach inherently preserves privacy, as nodes determine their status using only distance vectors without reconstructing the global topology. 
Future work will focus on extending this framework to asynchronous communication environments and investigating its applicability to directed graphs.

% \begin{ack}
% Place acknowledgments here.
% \end{ack}

\bibliography{ifacconf}             % bib file to produce the bibliography
                                                     % with bibtex (preferred)

\appendix

\section{Design Rationale for Local State Update on Edge Change}
\label{appendix:Design Rationale}
% The dynamic update mechanism is designed to correctly and efficiently propagate the effects of topological changes across the network, ensuring that all nodes eventually converge to a consistent and accurate view of the new connectivity structure. The core intuition behind the design choices is rooted in the relationship between shortest-path distances and reachability in the context of edge modifications.

The state update \eqref{eq:immediate change} at the incident nodes $i$ and $j$ immediately after a change is a conservative invalidation mechanism. It is designed to immediately reset any state $x_{ik}$ that is potentially invalidated by the modification of edge $(i,j)$, while preserving any state that is unaffected. The decision relies on the distance difference $\Delta_{ik}^j$, which evaluates the topological relationship between nodes $k$, $i$, and $j$.
% based on a conservative principle from shortest-path properties. 

First, we consider the case where $\Delta_{ik}^j \leq 0$, meaning node $k$ is closer to (or equidistant from) $i$ than to $j$. 
If an edge is added (see Fig.~\ref{fig:sub_a}), a new potential path $(k, \ldots, j, i)$ is created with length $d_{jk} + 1$. From the case condition, we have $d_{ik} \leq d_{jk} < d_{jk} + 1$. This means the new path is strictly longer than the existing shortest path(s).
If an edge is deleted (see Fig.~\ref{fig:sub_b}), we analyze if any shortest path from $k$ to $i$, $P_{ki}$, could have used the edge $(j,i)$ as its final hop. If it did, its length would be $d_{ik} = d_{jk} + 1$. This directly contradicts the case condition $d_{ik} \leq d_{jk}$. Therefore, no shortest path from $k$ to $i$ utilized the edge $(i,j)$, and its removal does not affect the shortest-path structure.
In both scenarios, the shortest-path structure (both distance and path set) from $k$ to $i$ is provably unaffected by the change. Therefore, The state $x_{ik}$ remains valid.

Second, we consider the case where $\Delta_{ik}^j > 0$, meaning node $k$ was closer to $j$ than to $i$.
If an edge is added (see Fig.~\ref{fig:sub_c}), a new path $(k, \ldots, j, i)$ (length $d_{jk} + 1$) is introduced. The new shortest path distance will be $\min(d_{ik}, d_{jk} + 1)$. Since $d_{ik} > d_{jk}$, it is possible that $d_{jk} + 1 \leq d_{ik}$. This means the shortest path distance may decrease, or new shortest paths may be formed.
If an edge is deleted (see Fig.~\ref{fig:sub_d}), the edge $(i,j)$ must have existed. By the triangle inequality, $d_{ik} \leq d_{jk} + 1$. Combining this with the case condition $d_{ik} > d_{jk}$ forces the exact equality $d_{ik} = d_{jk} + 1$. This equality proves that the path $(k, \ldots, j, i)$ (where $(k, \ldots, j)$ is a shortest path) is a shortest path from $k$ to $i$. Deleting the edge $(i,j)$ breaks this path, thus invalidating $x_{ik}$. 
In both situations, the shortest-path structure that $x_{ik}$ was based on is now invalid. The state $x_{ik}$ is therefore conservatively reset to $0$, forcing a re-evaluation of this reachability through subsequent consensus iterations.

\section{PROOF OF LEMMA~\ref{lemma:conservative_aggregation}}
\label{appendix:lemma:conservative_aggregation}
Consider the $k$-th component of the reachability vectors, corresponding to destination $k$. First, Equation \eqref{eq:Vee-Intra-Flag} computes the component $(x_{\phi})_k = \max_{j \in \mathcal{S}(\phi)} x_{jk}[t-1]$. By the definition of logical disjunction, $(x_{\phi})_k$ evaluates to 1 if and only if the set $\{j \in \mathcal{S}(\phi) \mid x_{jk}[t-1]=1\}$ is non-empty. This implies that for the specific event $\phi$, at least one neighbor $j$ among those propagating $\phi$ is capable of reaching $k$. Second, Equation \eqref{eq:Wedge-Inter-Flag} computes $\widetilde{x}_{ik}$ using a logical conjunction ($\bigwedge$) over all active flags $\phi \in \Phi_i[t]$. By the definition of logical conjunction, $\widetilde{x}_{ik}$ remains 1 if and only if $(x_{\phi})_k = 1$ for every $\phi$. Combining these two steps, the condition $\widetilde{x}_{ik}[t]=1$ holds if and only if for every detected event $\phi$, there is a neighbor $j$ such that $x_{jk}[t-1]=1$.\qed

\section{PROOF OF THEOREM~\ref{thm:convergence}}
\label{appendix:thm:convergence}
To prove Theorem~\ref{thm:convergence}, we analyze the evolution of the states $(x[t], d[t])$ within the static interval $[T, T+2n]$.

First, we address the extinction of change flags within the interval $(T, T+n]$.
Let $D$ be the diameter of the network, bounded by $n$. Any change flag $\phi$ existing in the network at time $T$ propagates one hop per time step via \eqref{eq:phi_tilde_update} and is recorded in the history set $\mathcal{H}$ via \eqref{eq:history_update}. Since $\mathcal{H}$ prevents reprocessing, every flag exists in the active set $\Phi$ of any node for exactly one time step. Consequently, by time $T_{purge} = T + n$, all flags generated at or before $T$ will have traversed the network and been purged from all active sets. 
% Thus, satisfies $\Phi_i[t] = \emptyset$ and operates under Case 3.

During the propagation, the algorithm ensures that all invalid distances are reset through local correction and consensus propagation. Consider a local edge change occurring at $t\le T$ that invalidates a path, the local state correction step detects this via the metric calculated via \eqref{eq:Delta} and accordingly resets the reachability state $x$ via \eqref{eq:immediate change}, which in turn forces the distance state $d$ to $\infty$.
For nodes not incident to the change, the invalidation propagates via flags.
Suppose the shortest path from $i$ to $k$ is affected by a topological event $\phi$ at or before $T$.
When the $\phi$ reaches node $i$ at time $t \le T_{purge}$, the neighbor $j$ on the affected path has already set $x_{jk}[t-1]=0$.
Consequently, the condition in Lemma~\ref{lemma:conservative_aggregation} is violated. The inter-flag aggregation \eqref{eq:Wedge-Inter-Flag} yields $\widetilde{x}_{ik}[t]=0$ and \eqref{eq:Final-Consensus} updates $\widehat{x}_{ik}[t] = \overline{x}_{ik}[t] \wedge 0 = 0$.
This $1 \to 0$ transition triggers the distance reset condition in \eqref{eq:distance-update-general}, forcing $\widehat{d}_{ik}[t] \to \infty$. 
% If 有本地的边的变化在t时刻，要做correction，具体对应x的修正{eq:immediate change}和d的，要说明修正也 ensures that all invalid distances are reset
Therefore, by time $T_{purge}$, all reachability states $x$ and distances $d$ corresponding to paths invalidated by prior events are correctly reset to $0$ and $\infty$, respectively.

% If a valid path exists in $\mathcal{G}_{final}$, the distance converges to the true shortest path length. 
Consider a node $i$ that has a valid path to $k$ exists, but currently $x_{ik}[t-1]=0, d_{ik}[t-1]=\infty$.
Any neighbor $j$ that has already processed the relevant flags is considered a historical source. According to \eqref{eq:Historical-Update}, if such a neighbor has a valid path ($x_{jk}[t-1]=1$), it drives $\overline{x}_{ik}[t] \to 1$.
Since no new active flags exist to block this update (i.e., $\widetilde{x}_{ik}[t]=1$), the final state becomes $x_{ik}[t]=1$.
This $0 \to 1$ transition triggers the distance update $d_{ik}[t] = 1 + \min_{j \in \mathcal{N}_i} d_{jk}[t-1]$ defined in Eq. \eqref{eq:distance-update-general}.
By induction, if a neighbor $j$ has converged to the true shortest path distance $L-1$ at time $t-1$, node $i$ computes $d_{ik}[t] = 1 + (L-1) = L$. Since no neighbor can offer a distance less than $L-1$, the value $d_{ik}[t]=L$ is correct.

Since the maximum shortest path length is bounded by $n$, all nodes converge to the correct distance $d^*$ within at most $n$ additional steps via \eqref{eq:distance} after $T_{purge}$. The total convergence time is the sum of the flag purge duration and the relaxation duration, bounded by $2n$. Therefore, after $2n$ rounds (i.e., at time $T + 2n$), the states $(x, d)$ are guaranteed to converge to the correct states of $\mathcal{G}[T]$.
\qed

\end{document}